\documentclass[preprint,12pt]{elsarticle}

\usepackage{amssymb}
\usepackage{amsmath}

\usepackage{graphicx}
\usepackage{subcaption}

\usepackage[table]{xcolor}

\usepackage{tabularx}
\usepackage{booktabs}
\usepackage{multirow}
\usepackage{array}
\usepackage{makecell}
\usepackage{adjustbox}
\usepackage{rotating}

\usepackage{comment}
\usepackage{multicol}
\usepackage[section]{placeins}

\usepackage{tikz}
\usepackage{tikz-cd}
\usetikzlibrary{arrows.meta,positioning,calc,backgrounds}

\usepackage{algorithm}
\usepackage{algpseudocode}

\usepackage[hidelinks]{hyperref}

\journal{Sustainable Energy, Grids and Networks}

\begin{document}

\begin{frontmatter}



\title{Hard-Constrained Probabilistic Factor Graph Neural Network for Distribution System State Estimation under Non-Gaussian Uncertainty}
\author[label1,label2,label3]{M. Furqan Azam}
\ead{furqan.azam@kuleuven.be}
\author[label1,label3]{Marta Vanin}
\author[label2,label3]{Chris Hermans}
\author[label1,label3]{Geert Deconinck}

\affiliation[label1]{organization={ELECTA, Department of Electrical Engineering, KU Leuven}, city={Heverlee},
  postcode={3001}, country={Belgium}}
\affiliation[label2]{organization={Flemish Institute for Technological Research (VITO)}, city={Mol}, postcode={2400},
  country={Belgium}}
\affiliation[label3]{organization={EnergyVille}, addressline={Thor Park}, city={Genk}, postcode={3600}, country={Belgium}}

\begin{abstract}
Robust and accurate state estimation is fundamental for the reliable operation and monitoring of active distribution networks. Conventional numerical estimators, such as weighted least squares, are computationally slower and often suffer from convergence issues in the presence of sparse measurements affected by non-Gaussian noise. Physics-informed neural networks have recently emerged as a promising alternative by incorporating physical principles through residual-based penalty terms in the objective, which can improve robustness to noise and computational efficiency. However, such penalty-based approaches do not guarantee strict enforcement of physical constraints during inference and provide limited support for principled uncertainty modeling. To address these limitations, we propose a novel Hard-Constrained, Physics-Informed Factor Graph Neural Network (HCP-PINN) that formulates distribution system state estimation as a probabilistic constrained estimation problem on a factor graph. The proposed approach explicitly models non-Gaussian (pseudo-)measurement uncertainty through flexible, closed-form likelihood-based loss functions. It further incorporates a differentiable optimization-based estimation layer that strictly enforces nonlinear equality constraints and measurement consistency, producing physically feasible state estimates during both training and inference. To the best of our knowledge, this is the first PINN-based DSSE framework to combine hard physical constraints with non-Gaussian uncertainty modeling for unbalanced three-phase networks. Numerical experiments demonstrate that the proposed method delivers more accurate and physically consistent state estimates than deep learning benchmarks and numerical estimators, while exhibiting greater robustness to network model errors and improved computational efficiency under realistic measurement scenarios.

\end{abstract}

%
\begin{graphicalabstract}
    \centering
    \includegraphics[
        width=1.1\textwidth,
        trim=0cm 0cm 0cm 0cm, 
        clip
    ]{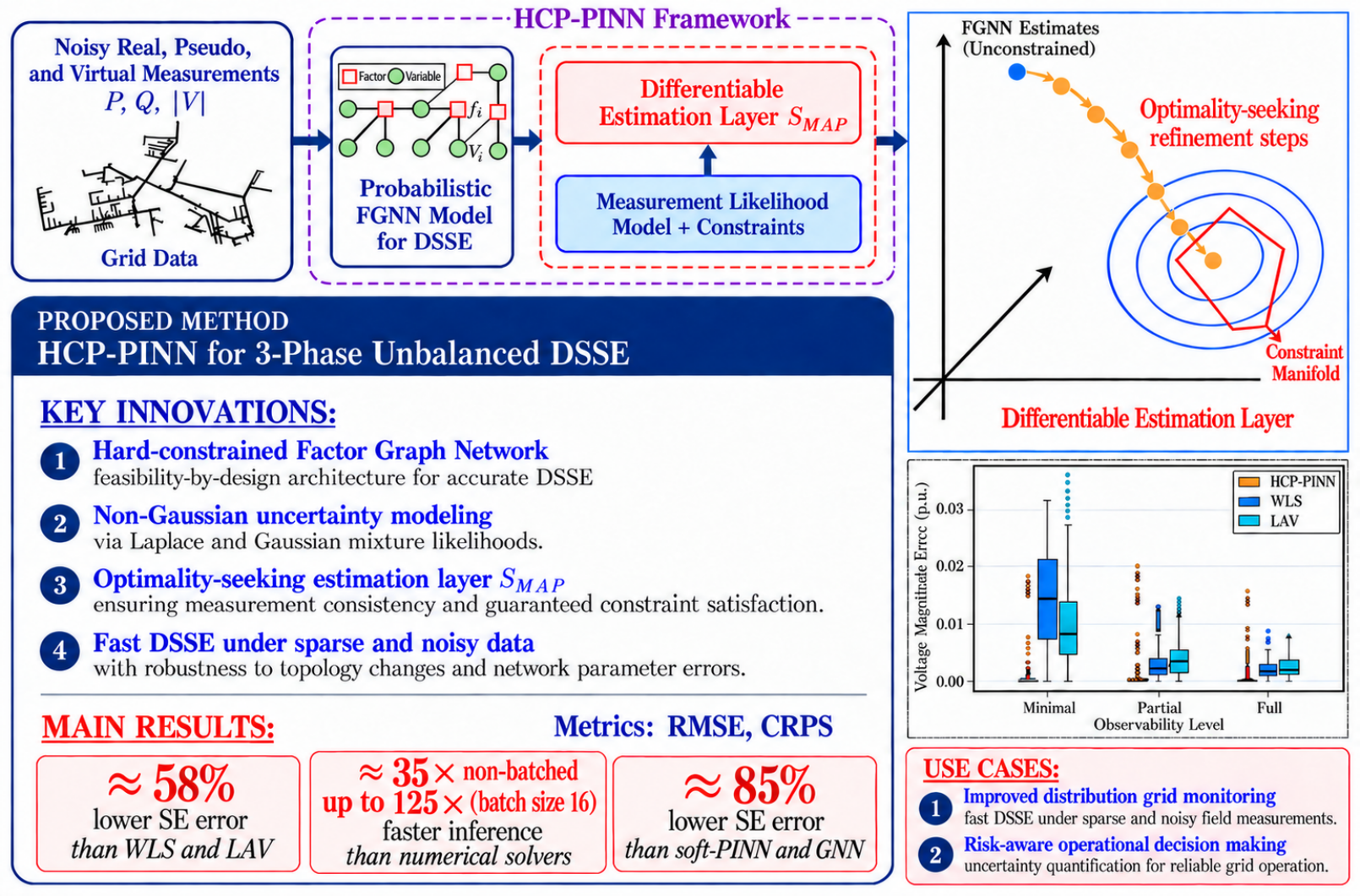}
\end{graphicalabstract}
%
\begin{highlights}
\item Introduces a novel hard-constrained probabilistic factor graph neural network framework, termed HCP-PINN, for fast and accurate DSSE under sparse and noisy measurement conditions, bad data, and topological changes.
\item Incorporates a novel differentiable estimation layer, \(\mathcal{S}_{\mathrm{MAP}}\), that strictly ensures physical consistency by performing optimality-seeking refinement steps on prior state estimates.
\item Supports continuous non-Gaussian uncertainty modeling through a generalized likelihood-based loss function, providing well-calibrated probabilistic state estimates.
\item Presents a geometric surrogate-gradient method to enable stable and computationally efficient model training without fully differentiating through the embedded \(\mathcal{S}_{\mathrm{MAP}}\) layer.
\item Provides a modular and computationally efficient framework for balanced and unbalanced DSSE that consistently outperforms numerical WLS, LAV, penalty-based PINNs, and purely data-driven deep learning approaches.
\end{highlights}


\end{frontmatter}




\section{Introduction}
\label{sec1}

\noindent\textit{\textbf{Motivation:}} Distribution networks (DNs) play an increasingly important role in the global energy transition. The rapid adoption of
rooftop photovoltaic (PV) systems, electric vehicles, and large loads such as heat pumps can strain existing electrical networks and cause congestion~\cite{wesseling_unbalanced}. Therefore, accurate, near-real-time network monitoring is required for reliable operation and advanced control, which is typically enabled by state estimation (SE) tools. SE is a statistical inference method that combines network models, including topology and line parameters, with a set of noisy field measurements to estimate the most likely system state. It is widely acknowledged that SE techniques developed for transmission systems are not directly applicable to DNs, which are characterized by large phase unbalances, radial or weakly meshed topologies and limited measurement coverage. These characteristics often result in low observability and poorly conditioned estimation problems, making conventional SE considerably more challenging in DNs~\cite{ahmad2018dsse_smartgrid}. Among the challenges that hinder the adoption of distribution system state estimation (DSSE) approaches in real-world DNs, the following are particularly important:

\begin{itemize}
  \item A scarcity of high-quality field measurements with low time resolution, resulting in limited network observability and measurement coverage.
  \item Measurement errors that are frequently non-Gaussian and heteroscedastic (and may be multimodal), making conventional weighted least squares (WLS) estimators statistically mismatched and potentially biased.
  \item The high computational cost of numerical estimators can hinder near-real-time or online monitoring in large-scale DNs, where system states must be estimated frequently.
\end{itemize}


Limited observability in DNs is commonly addressed by introducing pseudo-measurements, i.e., estimates derived from
statistical information or forecasts of non-metered nodal power injections; however, their inherent low accuracy can strongly
affect DSSE performance~\cite{angioni2016impact}. In standard SE practice, measurement and pseudo-measurement errors are often assumed to follow zero-mean Gaussian distributions, motivating the use of WLS-based estimators~\cite{angioni2016realtime,kotha2023wams_robustlwls}. However, residential demand and renewable generation forecast errors are frequently better described by non-Gaussian model such as Laplace, Student-$t$, or Gaussian-mixture model (GMM) distributions~\cite{singh2010gmm_pseudomeas,heunis2002probabilistic_loads,valverde2013stochastic_monitoring}. This motivates the development of advanced DSSE formulations that can flexibly accommodate diverse uncertainty models while remaining computationally efficient for large-scale deployment, rather than assuming Gaussian measurement errors.
Deep learning (DL) approaches, such as graph neural networks (GNNs), can enable faster DSSE while improving robustness to noise and missing data across diverse operating conditions~\cite{ringsquandl2021relationalbias,kundacina2022gnn_se}. However, purely data-driven GNNs do not strictly enforce physical constraints, such as nodal power balance, and may therefore produce physically inconsistent state estimates, which limits their reliability for safety-critical power system applications.

\noindent\textit{\textbf{Related Work:}} Empirical studies indicate that field measurement errors can exhibit heavy-tailed and non-Gaussian characteristics. Similar non-Gaussian characteristics have also been observed in phasor measurement unit (PMU) data in transmission system state estimation studies~\cite{wang2018gaussian,huang2021nonGaussian}. Non-Gaussian distributions, such as Laplace or GMMs, often provide a better fit, particularly at buses with prosumers or mixed load–generation profiles~\cite{singh2010statistical_load_gmm}. WLS corresponds to the maximum-likelihood estimator when the measurement errors are Gaussian with known error covariance. However, heavy-tailed, heteroscedastic, or outlier-contaminated measurements can violate this assumption and degrade estimation performance. This has motivated the development of robust estimators that reduce sensitivity to outliers and non-Gaussian errors, including the least absolute value (LAV) estimator~\cite{gol2015hybrid_pmu} and Huber-based M-estimators~\cite{zhao2016sparse_vs_gmle}. The Weighted Least Absolute Value (WLAV) estimator is a widely used robust SE method that limits the influence of outliers by minimizing a weighted $\ell_1$ norm of the measurement residuals~\cite{kotiuga1982bad_data_wlav,lin2018robust_paramerrors}. Unlike WLS, it remains effective under non-Gaussian or unknown noise distributions.


For exact modeling of non-Gaussian noise, Vanin et al.~\cite{vanin2023exact} proposed a general framework that supports arbitrary continuous (pseudo-)measurement uncertainty models by formulating DSSE as a constrained optimization problem. However, such approaches typically require the measurement error distribution to be known \emph{a priori}, which is not always practical. Related Bayesian-based DSSE approaches that incorporate more flexible noise models have also been explored; however, they are typically more computationally demanding than WLS- or LAV-based methods and have primarily been evaluated on smaller test cases~\cite{pegoraro2017bayesian_nonGaussian_dsse}. Kalman filter--based methods have been widely studied for dynamic SE, particularly in transmission systems~\cite{liu2020comparisons_kf_dse, zhao2019ukf}. However, the performance of these approaches can degrade in the presence of strong nonlinearities and numerical ill-conditioning induced by sparse sensing and noisy pseudo-measurements in real DNs. Earlier studies ~\cite{huang2015evaluation} report no significant advantage of dynamic over static DSSE in distribution systems, largely due to the limited availability of $\mu$PMUs and the heavy reliance on pseudo-measurements. Therefore, in this paper, we focus on addressing the static DSSE problem. To handle unknown or mismatched noise statistics, Cheng et al.~\cite{cheng2021glm} proposed an adaptive SE technique based on Gaussian–Laplacian mixture models for modeling measurement error distributions. 
%
%
Distributed SE has been widely studied to address the computational and scalability challenges of centralized SE by partitioning buses and measurements across multiple control areas and coordinating the solutions of local estimation problems~\cite{Chen2021ADR}. More recently, factor graph-based formulations have been explored as a scalable and modular way of solving SE that supports distributed inference via message passing~\cite{cosovic2016dc,hu2011bp_dsse}. However, most practical implementations rely on simplified Gaussian error models and linearized message updates, which can be sensitive to bad data, heavy-tailed noise, and pseudo-measurements. Moreover, many studies assume large PMU deployments that are less common in DNs.


Recently, DL-based methods, particularly GNNs, have shown promise for fast and robust DSSE by exploiting grid topology under noisy measurements~\cite{moshtagh2025topologyaware_gnn_se,zargar2020multiarea_pmu_dsse,lin2022elegnn}. These methods learn a surrogate mapping from sparse measurements and topology to system states, avoiding repeated nonlinear solves at each timestep. Their data-driven formulation can also provide robustness to uncertain or partially inaccurate network parameters~\cite{zhang2025explainable_mfbnn_dsse}. However, most existing GNN-based DSSE approaches rely on supervised learning and thus require extensive labeled datasets of ground truth system states, which are typically unavailable in practical DN scenarios. In addition, their physics-agnostic nature can lead to physically inconsistent estimates that can violate power balance constraints, particularly under out-of-distribution operating scenarios. Furthermore, a key representational limitation of standard GNNs is their reliance on local pairwise message passing, which may be insufficient to capture the higher-order multivariate dependencies induced by the power flow physics. Donon et al.~\cite{donon2020deep_statistical_solvers} highlight similar shortcomings of standard GNN models in capturing the structural complexity of power grids, which can result in information loss. In DSSE, for instance, the voltage at a bus can depend jointly on multiple neighboring bus voltages, power injections or branch flows, and line parameters. Collapsing these interactions into indiscriminate feature aggregations during GNN message passing may dilute structured nonlinear relationships that depend jointly on several variables through the power flow equations.

Motivated by the lack of physical feasibility guarantees in purely data-driven approaches, physics-informed neural network (PINN) approaches have recently gained traction for performing DSSE~\cite{zamzam2020physicsaware_dsse,cao2024piglb_dsse,authier2024graphyr_dyr,azam2024lv_pf_pignn}. PINNs incorporate physical laws and constraints (e.g., Kirchhoff’s law or power balance) into the learning process to improve robustness and generalization, especially under noisy and sparse measurement scenarios. For instance, authors in~\cite{dejongh2022physicsinformed_gdl} propose a physics-guided geometric deep learning model that improves SE accuracy under noisy data and unseen grid topologies. Similarly, Habib et al.~\cite{habib2023deep_statistical_solver_dsse} introduced a weakly supervised learning strategy in which a neural network is trained with a WLS-style objective function, allowing it to learn from both smart meter measurements and low-fidelity pseudo-measurements.

However, existing PINN-based DSSE methods typically enforce physical constraints only \emph{softly} by incorporating physics residuals as weighted penalty terms in the training objective. While this promotes physics consistency, it does not guarantee exact constraint satisfaction during inference and remains sensitive to penalty-weight tuning. Poorly tuned weights can lead to persistent constraint violations and reduced generalization under challenging or out-of-distribution conditions, such as unseen feeder topologies. In our previous work~\cite{azam2026physics}, we proposed a physics-informed factor graph neural network for DSSE that enforces power balance constraints during training through an augmented Lagrangian formulation with adaptive weighting. While this improves constraint satisfaction and reduces reliance on manual penalty-weight tuning, exact satisfaction of the nonlinear power balance constraints is still not guaranteed during inference.

To overcome the limitations of soft-constrained PINNs, hard-constrained neural network frameworks have been explored in other domains~\cite{grontas2025pinet,irawan2026hardconstrained_pinns}. However, to the best of our knowledge, their application to SE remains largely unexplored, particularly for unbalanced three-phase networks. Such approaches typically embed system constraints directly into the network architecture through projection or differentiable optimization layers that map the neural network output onto the feasible set. While this ensures constraint satisfaction, it does not necessarily optimize the underlying nonlinear estimation objective. This distinction is particularly important in DSSE, where noisy and incomplete measurements require the state to be inferred by jointly considering measurement consistency and network physics. Hence, the embedded optimization layer should go beyond feasibility restoration and solve a constrained SE subproblem that simultaneously enforces the physical constraints and minimizes the measurement residuals.
  
Other studies \cite{okhuegbeMachineLearningInitializer2024} use neural networks to warm-start numerical solvers, thereby improving convergence and reducing overall computation time. However, the downstream solver is non-differentiable and excluded from backpropagation, such that the neural network is trained without direct feedback from the final solver solution. Consequently, poor generalization of the neural network initializer under unseen conditions can diminish the computational benefit of the warm start and, in some cases, lead to solver divergence. End-to-end model training with an embedded optimization solver requires differentiating through the iterative solver during backpropagation, which can be computationally expensive and numerically fragile for large networks with sparse and noisy measurements. Existing approaches typically rely on either \emph{explicit (unrolled) differentiation}, which backpropagates through the full sequence of solver iterations~\cite{xie2025neural}, or \emph{implicit differentiation}, which differentiates the first-order optimality conditions, often expressed through the Karush--Kuhn--Tucker (KKT) system, of the embedded optimization problem~\cite{donti2021dc3}. Both approaches face challenges in nonlinear problems: unrolling or explicit differentiation incurs high computational and memory costs, whereas implicit differentiation relies on reliable convergence to a well-defined solution satisfying the fixed-point or optimality conditions, which can be difficult to achieve under sparse and noisy data. Moreover, implicit differentiation is applicable only when the neural network outputs appear in the KKT or optimality conditions defining the final solution, which is generally not the case when they are used solely as initialization points. To address the computational burden of unrolling or explicit differentiation, surrogate-gradient methods provide an efficient alternative for backpropagating through optimization layers~\cite{rashwan2025enforcing,gould2023gradient}. A common approach is the straight-through estimator (STE), which passes gradients through the embedded layer without differentiating its underlying operation~\cite{Bengio2013EstimatingOP}. However, this approximation ignores the geometry of the constrained transformation and may produce gradients that are poorly aligned with the feasible solution manifold. This work therefore proposes a custom \emph{geometric surrogate-gradient} method that accounts for the geometry of the constraint manifold while avoiding the computational cost and the memory overhead of solver unrolling.

Most existing PINN-based formulations, whether soft- or hard-constrained, remain largely deterministic and provide limited support for predictive uncertainty. In DSSE, however, uncertainty due to heterogeneous sensor noise, limited observability, and grid model errors can directly affect the reliability of downstream operational decisions. This underscores the need for probabilistic DSSE frameworks that move beyond point estimation by explicitly propagating measurement uncertainty through the nonlinear power flow equations to the final state outputs. This work enables uncertainty quantification for DSSE by learning the parameters of a predictive distribution over the system state and training the model with likelihood-based objectives. This probabilistic formulation allows the model to capture non-Gaussian uncertainty while preserving strict physical consistency and constraint satisfaction through the embedded optimization layer.




\noindent\textit{\textbf{Contributions:}} This work proposes HCP-PINN, a \textit{Hard-Constrained, Probabilistic, Physics-Informed Factor Graph Neural Network} framework for accurate and computationally efficient DSSE. HCP-PINN formulates DSSE as a constrained estimation problem on a factor graph, where each measurement and physical relation is modeled as a local multivariate factor node. This representation preserves higher-order state-measurement dependencies and enables structured inference under noisy and sparse measurement coverage. The model is trained by minimizing a generalized negative log-likelihood (NLL) objective. During the forward pass, an embedded differentiable estimation layer computes a constrained \emph{maximum a posteriori} (MAP) estimate that ensures physical consistency by enforcing the equality constraints. The proposed estimator is \emph{feasible-by-design}, meaning that feasibility is guaranteed by construction rather than achieved through soft penalties or post-hoc correction. The proposed framework aims to achieve state-of-the-art DSSE accuracy with guaranteed constraint satisfaction, while supporting non-Gaussian uncertainty modeling, improving scalability, reducing observability requirements relative to numerical WLS and robust LAV estimators, and enhancing robustness to network parameter errors and future grid upgrades. To the best of the authors' knowledge, this paper presents the \textit{first implementation of a hard-constrained probabilistic PINN framework for DSSE in both balanced and unbalanced networks under realistic non-Gaussian measurement uncertainty}, enabled by the following key contributions:
\begin{itemize}
  \item A probabilistic factor graph neural network (FGNN) model with variable and factor nodes, enabling structured inference that preserves higher-order dependencies and supports principled uncertainty quantification under noisy, low observability settings.

  \item A differentiable, optimality-seeking estimation layer that solves a constrained MAP subproblem by combining learned prior uncertainty with measurement models, while strictly enforcing power balance constraints on zero-injection buses.

  \item Non-Gaussian uncertainty modeling via a generalized training objective that minimizes a closed-form NLL loss for a broad class of parametric distributions, thereby enabling flexible uncertainty quantification.

  \item An efficient, GPU-friendly geometric surrogate-gradient method that avoids unrolling iterative solvers during backpropagation, reducing the computational cost and memory overhead typical of existing projection-based approaches.
    
\end{itemize}

The proposed model is evaluated on both single-phase equivalent balanced  and three-phase unbalanced test networks of varying sizes, under partial observability conditions, non-Gaussian noise settings, feasibility, network parameter errors, topological reconfigurations and out-of-distribution loading scenarios. The experimental case studies demonstrate that the proposed HCP-PINN consistently outperforms conventional WLS, robust LAV estimators, soft-constrained PINN methods, and purely data-driven GNN approaches in terms of estimation accuracy, constraint satisfaction, while also providing a modular architecture and high computational efficiency.


The remainder of this paper is organized as follows: Section~\ref{sec2} presents the constrained DSSE problem formulation, followed by the proposed methodology in Section~\ref{sec3}. The experimental setup is described in Section~\ref{sec4}, while the numerical results are reported and discussed in Section~\ref{sec5}. Finally, Section~\ref{sec6} concludes the paper.


\section{Problem Formulation}
\label{sec2}
DSSE estimates the most likely system state from sparse and noisy field measurements by combining the network model with the corresponding nonlinear measurement equations~\cite{abur2004power}. The relationship between the system state and the available measurements is
described generally by a nonlinear measurement function $\mathbf{h}(\mathbf{x})$ as:
\begin{equation}
  \mathbf{z}=\mathbf{h}(\mathbf{x})+\boldsymbol{\varepsilon},
  \label{eq:measurement}
\end{equation}
where $\mathbf{x}\in\mathbb{R}^{2n-1}$ is the state vector collecting bus voltage magnitudes and phase angles, $\mathbf{x}=[\mathbf{V}^\top,\boldsymbol{\theta}^\top]^\top$, for a system with $n$ buses, with the slack bus angle fixed to $\theta_{\mathrm{slack}}=0$. The noise vector $\boldsymbol{\varepsilon}$ represents random measurement errors characterized by an assumed probability distribution. The measurement vector $\mathbf{z}\in\mathbb{R}^{|\mathcal{M}\cup\mathcal{P}|}$ stacks real measurements $\mathcal{M}$ (e.g., voltage magnitudes, power injections, and line flows from power meters) together with pseudo-measurements $\mathcal{P}$, i.e., estimated values derived from historical data to ensure measurement redundancy. The nonlinear measurement function $\mathbf{h}(\cdot)$, derived from the AC power flow equations, maps the state $\mathbf{x}=(\mathbf{V},\boldsymbol{\theta})$ to the measurement space by relating estimated voltage magnitudes and phase angles to all network quantities such as branch power flows $P_{ij}$ and bus power injections $P_i$. For readability and notational clarity, the following equations are expressed using a single-phase equivalent representation. In particular, for bus $i\in\mathcal{N}$, the measurement equations for voltage magnitude and phase angle are:
\begin{subequations}\label{eq:meas_explicit}
  \begin{align}
    z_{V_i} &= V_i + \varepsilon_{V_i}, \\
    z_{\theta_i} &= \theta_i + \varepsilon_{\theta_i},
  \end{align}
\end{subequations}
For power flowing from bus $i$ to bus $j$ via branch $(i,j)\in\mathcal{E}$,
\begin{subequations}\label{eq:meas_flows}
  \begin{align}
    z_{P_{ij}} &= V_i^2\,(g_{ij}+g_i^{\mathrm{s}}) - V_iV_j\!\left(g_{ij}\cos\theta_{ij}+b_{ij}\sin\theta_{ij}\right) + \varepsilon_{P_{ij}}, \\
    z_{Q_{ij}} &= -V_i^2\,(b_{ij}+b_i^{\mathrm{s}}) - V_iV_j\!\left(g_{ij}\sin\theta_{ij}-b_{ij}\cos\theta_{ij}\right) + \varepsilon_{Q_{ij}},
  \end{align}
\end{subequations}
where the line parameters consist of branch conductance $g_{ij}$, branch susceptance $b_{ij}$, and shunt elements (i.e. conductance $g_i^{\mathrm{s}}$ and susceptance $b_i^{\mathrm{s}}$). Measurement equations for the bus injections $P_i$ and $Q_i$ can be written as:
\begin{subequations}\label{eq:meas_inj}
  \begin{align}
    z_{P_i} &= V_i \sum_{j\in\mathcal{N}(i)} V_j\!\left(g_{ij}\cos\theta_{ij}+b_{ij}\sin\theta_{ij}\right) + \varepsilon_{P_i}, \\
    z_{Q_i} &= V_i \sum_{j\in\mathcal{N}(i)} V_j\!\left(g_{ij}\sin\theta_{ij}-b_{ij}\cos\theta_{ij}\right) + \varepsilon_{Q_i},
  \end{align}
\end{subequations}
where $\theta_{ij} = \theta_i - \theta_j$ and $\mathcal{N}(i)$ denotes the set of buses adjacent and connected to bus $i$. Each of the above measurement equations defines the corresponding entry of the nonlinear measurement function $\mathbf{h}(\cdot)$. Stacking them yields the compact vector form in \eqref{eq:measurement}. The corresponding residual vector is then defined as:
\begin{equation}
  \mathbf{r}(\mathbf{x}) := \mathbf{z}-\mathbf{h}(\mathbf{x}), \qquad r_j(\mathbf{x}) := z_j-h_j(\mathbf{x}).
  \label{eq:residual}
\end{equation}

Assuming that individual measurements are conditionally independent given the system state, the joint likelihood is given by:
\begin{equation}
    p(\mathbf{z}\mid\mathbf{x})
    =
    \prod_{k\in\mathcal{M}\cup\mathcal{P}}
    p_k\!\left(r_k(\mathbf{x})\right),
    \label{eq:likelihood}
\end{equation}
where $p_k(\cdot)$ denotes the probability density associated with
measurement error $k$. The SE can then be solved as a maximum likelihood estimation (MLE) problem. Maximizing
\eqref{eq:likelihood} is equivalent to minimizing the corresponding negative log-likelihood (NLL):
\begin{equation}
    J_{\mathrm{MLE}}(\mathbf{x})
    :=
    -\log p(\mathbf{z}\mid\mathbf{x})
    =
    \sum_{k\in\mathcal{M}\cup\mathcal{P}}
    \ell_k\!\left(r_k(\mathbf{x})\right),
    \label{eq:mle_obj}
\end{equation}
where
$\ell_k(\cdot):=-\log p_k(\cdot)$ denotes the corresponding NLL term.
The choice of $\ell_k$ determines the resulting estimator. When prior information about the system state is available, combining this likelihood with the state prior yields the MAP estimator. From Bayes' rule,
\begin{equation}
    p(\mathbf{x}\mid\mathbf{z})
    \propto
    p(\mathbf{z}\mid\mathbf{x})\,p(\mathbf{x}),
    \label{eq:bayes_rule}
\end{equation}
where \(p(\mathbf{x})\) denotes the prior distribution of the state variables.
Taking the negative logarithm and omitting terms independent of
\(\mathbf{x}\), the MAP objective is obtained by augmenting the MLE objective
with the negative log-prior:
\begin{equation}
    J_{\mathrm{MAP}}(\mathbf{x})
    :=
    J_{\mathrm{MLE}}(\mathbf{x})
    -
    \log p(\mathbf{x})
    =
    \sum_{k\in\mathcal{M}\cup\mathcal{P}}
    \ell_k\!\left(r_k(\mathbf{x})\right)
    -
    \log p(\mathbf{x}).
    \label{eq:map_objective}
\end{equation}
The MAP estimate is therefore given by
\begin{equation}
    \hat{\mathbf{x}}_{\mathrm{MAP}}
    =
    \arg\min_{\mathbf{x}}
    J_{\mathrm{MAP}}(\mathbf{x}).
\end{equation}
Exact zero-injection conditions can then be enforced as equality constraints $\mathbf{c}(\mathbf{x})$, yielding the following constrained MAP formulation:
\begin{equation}
    \arg\min_{\mathbf{x}}\;
    J_{\mathrm{MAP}}(\mathbf{x})
    \quad \text{subject to} \quad
    \mathbf{c}(\mathbf{x})=\mathbf{0}.
    \label{eq:constrained_map}
\end{equation}
This constrained MAP formulation forms the basis of the HCP-PINN framework developed in this work, where the measurement likelihood captures measurement uncertainty and the equality constraints enforce exact physical consistency. Under zero-mean Gaussian noise with known measurement variances, \eqref{eq:mle_obj} reduces to the WLS objective, whereas under zero-mean Laplace noise, it yields a LAV objective. Conventional WLS-based DSSE is typically solved iteratively using Gauss--Newton-type algorithms, which repeatedly linearize the nonlinear measurement model and update the system state. Under sparse and noisy measurements, however, the resulting estimation problem can become ill-conditioned and computationally demanding~\cite{abur2004power,primadianto2017review}.


\textit{\textbf{Factor graph equivalence}:} The factorization of the posterior distribution in \eqref{eq:bayes_rule} can be exploited to construct a structured \emph{factor graph} representation of DSSE problem. Specifically, each likelihood term $p_k(z_k \mid \mathbf{x})$ depends only on the subset of state variables involved in the corresponding measurement function $h_k(\mathbf{x})$, while prior information can be represented through additional factors. This dependency structure induces a bipartite factor graph, where variable nodes correspond to bus states and factor nodes encode individual measurement likelihoods, prior information, or physical constraints. Each measurement factor connects only to the state variables appearing in $h_k(\mathbf{x})$, resulting in a sparse, modular graph structure. 
This factor graph view of the MAP estimation problem naturally motivates message-passing formulations of DSSE and provides the foundation for the probabilistic FGNN architecture used in the proposed HCP-PINN framework.

\section{Proposed DSSE Methodology}
\label{sec3}

\begin{figure}[!t]
    \centering
    \includegraphics[
        width=1.05\linewidth,
        trim=20pt 0pt 10pt 0pt,
        clip
    ]{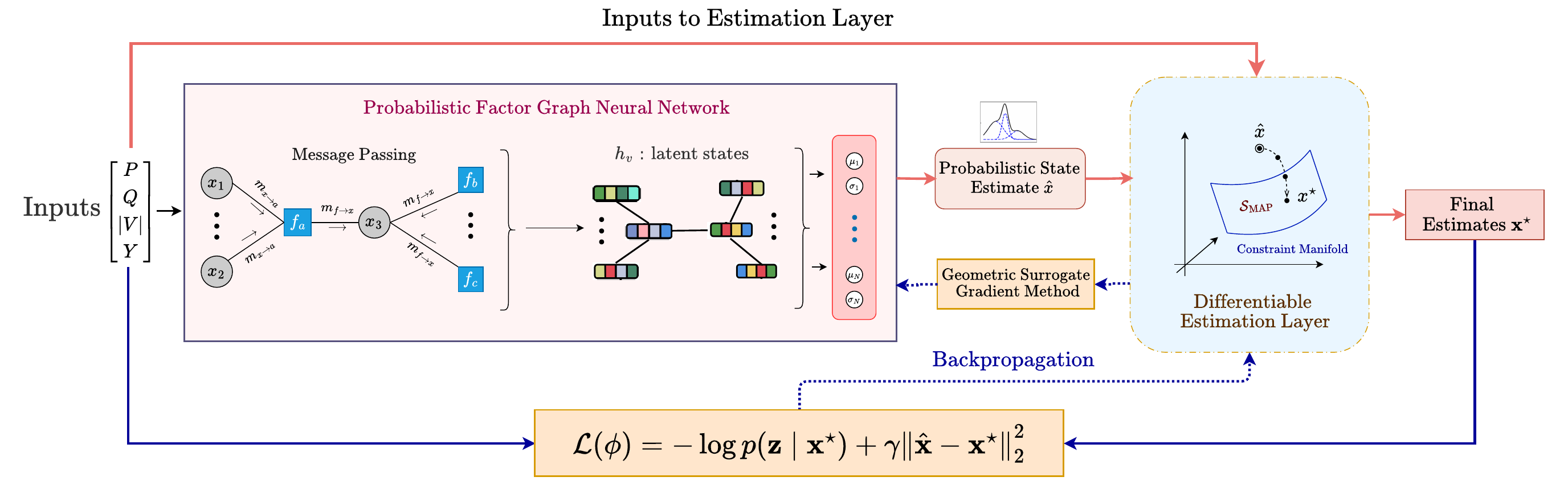}
    \caption{Overview of the proposed \textsc{HCP-PINN} framework. Input
    measurements and the network topology are processed by a probabilistic FGNN
    to produce a prior state estimate $\hat{x}$. The estimation
    layer $\mathcal{S}_{\mathrm{MAP}}$ combines this prior with the measurement
    likelihood while enforcing the equality constraints, yielding the refined
    state estimate $x^\star$.}
    \label{fig:proposed_framework}
\end{figure}

The proposed HCP-PINN framework, illustrated in
Fig.~\ref{fig:proposed_framework}, combines a probabilistic FGNN with a differentiable MAP estimation layer, $\mathcal{S}_{\mathrm{MAP}}$. The FGNN processes noisy measurements through factor graph message passing to parameterize the state distribution, while the $\mathcal{S}_{\mathrm{MAP}}$ layer refines the estimate using the measurement likelihood and explicit equality constraints. The framework is trained end-to-end with a likelihood-based loss function and a geometric surrogate-gradient method for efficient backpropagation through the $\mathcal{S}_{\mathrm{MAP}}$ layer.
The proposed framework employs the constrained MAP formulation in
\eqref{eq:constrained_map}, where the measurement likelihood is combined with
a learned prior $p_{\phi}(\mathbf{x})$ parameterized by the probabilistic
FGNN. The embedded estimation layer then solves:
\begin{equation}
    \mathbf{x}^{\star}(\mathbf{z};\phi)
    =
    \mathcal{S}_{\mathrm{MAP}}(\mathbf{z};\phi)
    :=
    \arg\min_{\mathbf{x}}\;
    J_{\mathrm{MAP}}(\mathbf{x};\mathbf{z},\phi)
    \quad
    \text{subject to}
    \quad
    \mathbf{c}(\mathbf{x})=\mathbf{0},
    \label{eq:single_map}
\end{equation}
where $\mathbf{c}(\mathbf{x})=\mathbf{0}$ denotes the power balance
constraints enforced at zero-injection buses. The FGNN parameterizes a learned prior $p_{\phi}(\mathbf{x}\mid\mathbf{z})$, whose negative log-probability acts as a probabilistic regularizer in the MAP objective $J_{\mathrm{MAP}}$ solved by the $\mathcal{S}_{\mathrm{MAP}}$ layer. For a Gaussian prior, the FGNN predicts the mean $\mu_{\phi}(\mathbf{z})$ and covariance $\Sigma_{\phi}(\mathbf{z})$. The resulting constrained MAP problem is solved by the $\mathcal{S}_{\mathrm{MAP}}$ layer using a fixed-iteration Gauss--Newton procedure with Levenberg--Marquardt damping~\cite{marquardt1963}. Starting from the FGNN initial prediction, the estimation layer iteratively minimizes the constrained objective while driving the solution toward the feasible manifold, thereby producing refined estimates that are both measurement-consistent and physically constrained.

The HCP-PINN framework is trained end-to-end using a generalized NLL objective augmented with an auxiliary consistency loss term:
\begin{equation}
\boxed{
\min_{\phi}\mathcal{L}(\phi)
=
-\log p_{\psi}\!\left(\mathbf{z}\mid\mathbf{x}^{\star}\right)
+
\gamma\left\|\hat{\mathbf{x}}-\mathbf{x}^{\star}\right\|_2^2
}
\label{eq:hcppinn_outer_loss}
\end{equation}
Here, $-\log p_{\psi}(\mathbf{z}\mid\mathbf{x}^{\star})$ denotes the  NLL term, selected according to the assumed measurement noise model, such as Gaussian, Laplace, or a GMM, while $\gamma>0$ weights the auxiliary consistency term that encourages the refined estimate $\mathbf{x}^{\star}$ to remain close to the initial FGNN prediction $\hat{\mathbf{x}}$. This limits large correction updates introduced by $\mathcal{S}_{\mathrm{MAP}}$, promoting better FGNN initialization and more stable convergence of the overall framework. The estimated state covariance is propagated through the nonlinear measurement function using the sampling-free, first-order delta method~\cite{vanDerVaart1998}, which linearizes $\mathbf{h}(\mathbf{x})$ around the refined estimate $\mathbf{x}^{\star}$ to obtain the approximate mean and covariance $(\hat{\mu},\hat{\Sigma})$ of the corresponding derived quantities, such as power injections and branch flows, for NLL evaluation. 
During model training, gradients are propagated through the $\mathcal{S}_{\mathrm{MAP}}$ layer using a geometric surrogate-gradient method, which avoids full unrolling of the iterative solver while preserving end-to-end differentiability. Next, we describe the key components of the proposed \textsc{HCP-PINN}; the probabilistic FGNN, the constrained estimation layer $\mathcal{S}_{\mathrm{MAP}}$, and the geometric surrogate-gradient method.

\subsection{Probabilistic Factor Graph Neural Network}
\label{subsec1}
\begin{figure}[!t]
  \centering
  \begin{minipage}[b]{0.48\columnwidth}
    \centering
    \includegraphics[trim=5cm 16.5cm 5cm 5cm,clip,width=\linewidth]{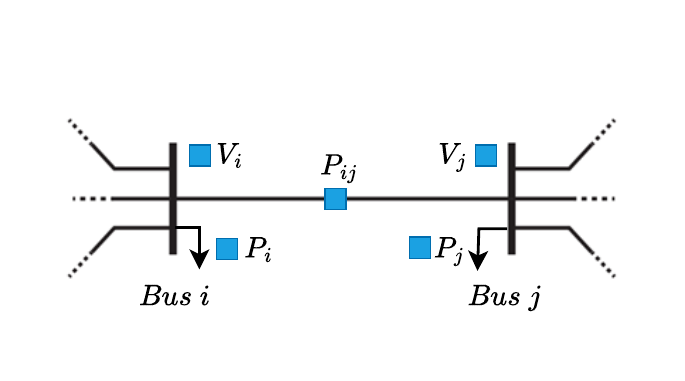}
    \vspace{3pt}
    {\small a) Physical two-bus system}
  \end{minipage}\hfill
  \begin{minipage}[b]{0.48\columnwidth}
    \centering
    \includegraphics[trim=1cm 0cm 0cm 0cm,clip,width=\linewidth]{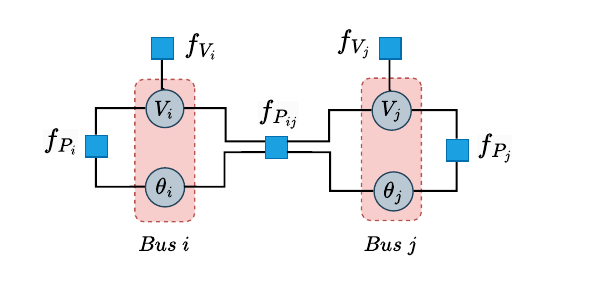}
    \vspace{3pt}
    {\small b) Factor graph representation}
  \end{minipage}
  \caption{Conversion of a 2-bus power system (left) into its factor graph representation (right). In the factor graph,
    voltages $V$ and angles $\theta$ are variable nodes (circles), while measurements are linked to factor nodes
  (squares).}
  \label{fig:factor_graph}
\end{figure}

The proposed HCP-PINN framework adopts the factor graph representation introduced in our previous work~\cite{azam2026physics} and extends it to a probabilistic, hard-constrained DSSE formulation for three-phase unbalanced DNs. As discussed earlier, standard GNNs perform pairwise message passing on the homogeneous bus-graph and may struggle to capture the higher-order dependencies arising from nonlinear power flow equations and multivariate relationships. In contrast, a factor graph representation models these multivariate interactions more naturally by explicitly expressing the joint likelihood as a product of local factors, thereby providing a more suitable graph structure for solving the DSSE problem. The proposed framework therefore employs a probabilistic FGNN architecture that performs message passing over measurement-likelihood factors and produces uncertainty-aware state estimates. Specifically, the power grid is represented as a bipartite factor graph $\mathcal{G}_f=(\mathcal{V}\cup\mathcal{F},\mathcal{E}_f)$, where variable nodes in $\mathcal{V}$ correspond to state variables (i.e., voltage magnitudes and phase angles) and factor nodes in $\mathcal{F}$ correspond to the individual measurement-likelihood terms. The edge set $\mathcal{E}_f$ is determined by the network model, measurement placement, and corresponding measurement equations, such that each factor node is connected only to the state variables involved in its associated measurement function. For architectural simplicity, constraint factors are omitted from the FGNN, and the corresponding equality constraints are instead enforced through the estimation layer $\mathcal{S}_{\mathrm{MAP}}$. Message passing is then performed over the resulting bipartite, heterogeneous graph while preserving the underlying physical topology.  Fig.~\ref{fig:factor_graph} illustrates how a simple 2-bus network is converted into its equivalent factor graph comprising variable nodes $(V,\theta)$ and factor nodes $f_{P_i}$ and $f_{P_{ij}}$.
The roles and input features of the FGNN graph nodes are defined as follows:
\begin{itemize}
  \item \textbf{Variable nodes:} represent the unknown state variables through latent embeddings, initialized using nominal state values
  (e.g., \( |V| \approx 1 \)~p.u. and \( \theta \approx 0 \)~rad).
    \item \textbf{Factor nodes:} represent local likelihood terms associated with individual measurements. Each factor encodes the measured value $z_i$, its uncertainty (e.g., variance $\sigma_i^2$), and the measurement type, and connects only to the state variables involved in the relation $z_i=h_i(\mathbf{x})+\varepsilon_i$.
\end{itemize}
This heterogeneous graph representation preserves the state--measurement
dependencies of the DSSE problem, accommodates diverse measurement types and higher-order interactions, and enables message passing that reflects the structure of the underlying estimation problem.
\begin{figure}[!t]
  \centering
  \begin{minipage}[b]{0.48\columnwidth}
    \centering
    \includegraphics[trim=6cm 21cm 6cm 1cm,clip,width=\linewidth]{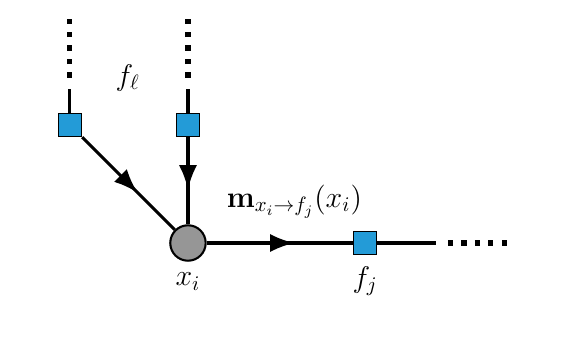}
    {\small a) V2F message passing}
  \end{minipage}\hfill
  \begin{minipage}[b]{0.48\columnwidth}
    \centering
    \includegraphics[trim=6cm 21cm 6cm 1cm,clip,width=\linewidth]{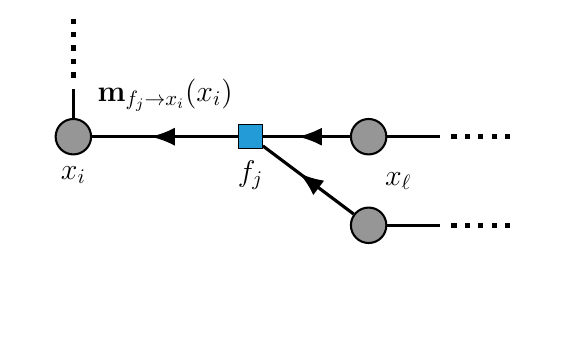}
    {\small b) F2V message passing}
  \end{minipage}
  \caption{Message passing on a factor graph. Left: V2F message, where a variable node $x_i$ aggregates incoming messages and sends its current state to a connected factor node $f_j$. Right: F2V message, where a factor node $f_j$ combines neighboring information with the associated measurement and sends an update to the target variable $x_i$.}
  \label{fig:message_passing}
\end{figure}
Inference on factor graphs is traditionally performed using the belief propagation (BP) algorithm, in which messages are iteratively exchanged between variable and factor nodes~\cite{NIPS2000_61b1fb3f}. However, analytical update rules in BP can be brittle, particularly under noisy and sparse measurement conditions, and may require many iterations to converge on large graphs. To address this limitation, the FGNN adopts a neural message passing approach that approximates BP. In the proposed model, the analytical update rules of classical BP are replaced by flexible neural network-based functions for message generation, aggregation, and state updates, as detailed in~\cite{azam2026physics}. During message passing, two types of messages are exchanged: variable-to-factor (V2F) messages and factor-to-variable (F2V) messages. In the V2F step, variable nodes send their current beliefs or intermediate state representations to the connected factor nodes. In the F2V step, factor nodes return measurement-based updates to the connected variable nodes, indicating how each local measurement influences the corresponding voltage magnitude or phase angle estimates while accounting for the measurement type and uncertainty.
The propagation of local V2F and F2V messages over the factor graph is illustrated in Fig.~\ref{fig:message_passing}. The resulting V2F and F2V message-update rules are defined as follows:
\begin{align}
    \mathbf{m}_{x_i \to f_j}^{(t)}
    &=
    \Phi_{\mathrm{V2F}}
    \left(
    \mathbf{h}_{x_i}^{(t)},
    \operatorname{AGG}
    \left\{
    \mathbf{m}_{f_\ell \to x_i}^{(t-1)}
    :
    f_\ell \in \mathrm{ne}(x_i)\setminus f_j
    \right\}
    \right), \\[1mm]
    \mathbf{m}_{f_j \to x_i}^{(t)}
    &=
    \Phi_{\mathrm{F2V}}
    \left(
    \mathbf{h}_{f_j}^{(t)},
    \operatorname{AGG}
    \left\{
    \mathbf{m}_{x_\ell \to f_j}^{(t)}
    :
    x_\ell \in \mathrm{ne}(f_j)\setminus x_i
    \right\},
    \mathbf{z}_j
    \right).
\end{align}

where $\mathbf{h}_{x_i}^{(t)}$ and $\mathbf{h}_{f_j}^{(t)}$ denote the latent embeddings of variable and factor nodes, respectively, while $\Phi_{\mathrm{V2F}}$ and $\Phi_{\mathrm{F2V}}$ are learnable multilayer perceptron (MLP)\footnote{An MLP is a feedforward neural network with multiple layers.}-based message functions. $\operatorname{AGG}$ denotes the aggregation operator that combines incoming messages from neighboring nodes into a single representation. The probabilistic FGNN reparameterizes the physical topology as a variable--factor computational graph, making the relationships between states and measurements explicit while preserving the underlying network topology. Moreover, sharing the MLP-based message and update functions across nodes of the same type enables the model to generalize across networks of varying sizes and topologies.

\subsection{Differentiable Estimation Layer $\mathcal{S}_{\mathrm{MAP}}$}
\label{subsec:prob_est_layer}

A key component of the \textsc{HCP-PINN} framework is the differentiable estimation layer $\mathcal{S}_{\mathrm{MAP}}$, which distinguishes this work from existing PINN-based DSSE approaches. It refines the unconstrained FGNN output into a constrained state estimate through a fixed number of optimality-seeking iterations that approximately solve the constrained \textit{maximum a posteriori} problem. The FGNN output provides both the initial state estimate and a learned prior. Within $\mathcal{S}_{\mathrm{MAP}}$, this prior is combined with the likelihood term to form the MAP objective, while equality constraints enforce active- and reactive-power balance at zero-injection bus phases. The prior thereby guides the refinement trajectory, optimality conditions, and resulting uncertainty estimate produced by the \(\mathcal{S}_{\mathrm{MAP}}\) layer. 
The probabilistic state estimate produced by the FGNN is represented as:
\begin{equation}
    \hat{\mathbf{x}}
    \sim
    \mathcal{D}\!\left(
        \boldsymbol{\mu}_{\mathrm{FGNN}},
        \boldsymbol{\Sigma}_{\mathrm{FGNN}}
    \right),
\end{equation}
where \(\mathcal{D}\) denotes the selected parametric distribution. The predicted mean initializes $\mathcal{S}_{\mathrm{MAP}}$, while the covariance determines the confidence assigned to the learned prior in the MAP objective. To relate the proposed  formulation to the conventional WLS-based SE case, first consider zero-mean
Gaussian measurement errors with covariance $\mathbf{R}$. The resulting
constrained MAP problem is given as:

\begin{figure}[!t]
  \centering
  \includegraphics[width=0.9\linewidth,trim=10pt 3pt 10pt 10pt,clip]{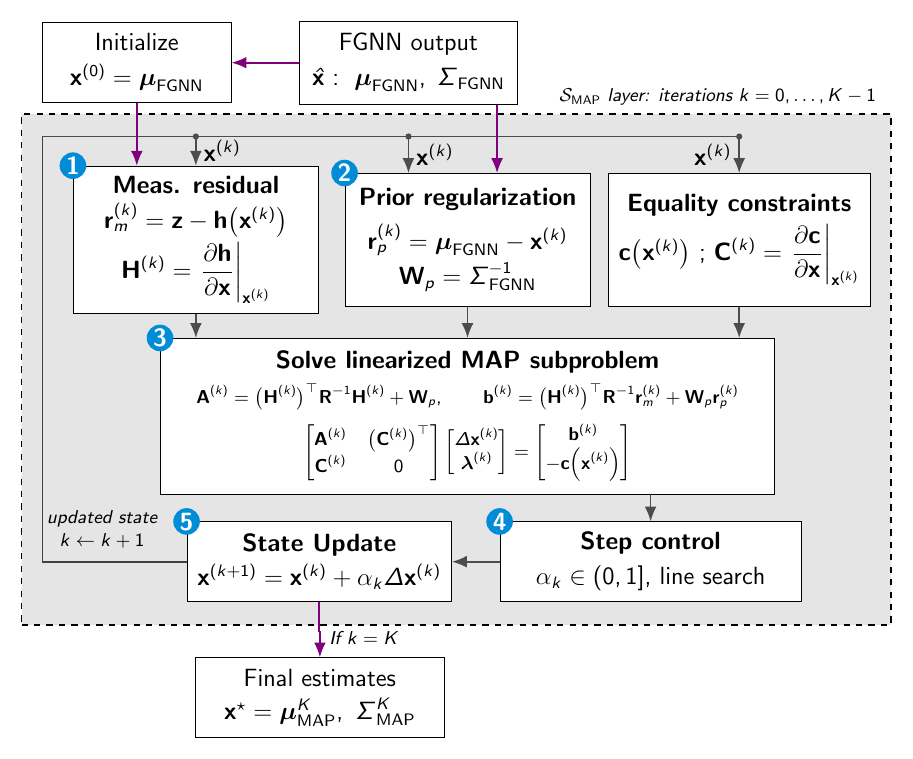}
    \caption{Differentiable $\mathcal{S}_{\mathrm{MAP}}$ layer. The FGNN prior initializes and regularizes $K$ damped Gauss--Newton updates, with hard equality constraints enforced throughout the refinement to obtain a physics-consistent MAP estimate and posterior covariance.}
  \label{fig:s_map_flow}
\end{figure}

\begin{equation}
\begin{aligned}
    \mathbf{x}_{\mathrm{MAP}}^{\star}
    =
    \underset{\mathbf{x}}{\arg\min}
    \quad &
    \frac{1}{2}
    \bigl(\mathbf{z}-\mathbf{h}(\mathbf{x})\bigr)^{\top}
    \mathbf{R}^{-1}
    \bigl(\mathbf{z}-\mathbf{h}(\mathbf{x})\bigr)
    \\
    &+
    \frac{1}{2}
    \bigl(\mathbf{x}-\boldsymbol{\mu}_{\mathrm{FGNN}}\bigr)^{\top}
    \boldsymbol{\Sigma}_{\mathrm{FGNN}}^{-1}
    \bigl(\mathbf{x}-\boldsymbol{\mu}_{\mathrm{FGNN}}\bigr)
    \\
    \text{subject to}
    \quad &
    \mathbf{c}(\mathbf{x})=\mathbf{0},
\end{aligned}
\label{eq:gaussian_constrained_map}
\end{equation}
Here, the first term penalizes the measurement residuals, with $\mathbf{R}^{-1}$ assigning larger weights to more reliable measurements. The second term regularizes the estimate toward $\boldsymbol{\mu}_{\mathrm{FGNN}}$, with prior precision $\boldsymbol{\Sigma}_{\mathrm{FGNN}}^{-1}$. Lower FGNN-predicted state uncertainty therefore results in stronger regularization toward the FGNN estimate, whereas higher state uncertainty allows larger corrections by $\mathcal{S}_{\mathrm{MAP}}$ layer.
Given the nonlinearity of the measurement function $\mathbf{h}(\mathbf{x})$ and equality constraints $\mathbf{c}(\mathbf{x})$, the constrained MAP problem in \eqref{eq:gaussian_constrained_map}  generally does not admit a closed-form solution. Instead, the proposed estimation layer approximately solves it using a fixed number $K$ of equality-constrained Gauss--Newton iterations. At iteration $k$, the measurement and constraint equations are linearized around the current estimate $\mathbf{x}^{(k)}$, and the state correction, together with the associated Lagrange multipliers, is obtained by solving the following KKT system:
\begin{equation}
\begin{bmatrix}
\mathbf{A}^{(k)} & \bigl(\mathbf{C}^{(k)}\bigr)^{\top} \\
\mathbf{C}^{(k)} & \mathbf{0}
\end{bmatrix}
\begin{bmatrix}
\Delta\mathbf{x}^{(k)} \\
\boldsymbol{\lambda}^{(k)}
\end{bmatrix}
=
\begin{bmatrix}
\mathbf{b}^{(k)} \\
-\mathbf{c}\!\left(\mathbf{x}^{(k)}\right)
\end{bmatrix}.
\label{eq:smap_kkt}
\end{equation}
where
\[
\begin{aligned}
\mathbf{A}^{(k)}
&=
\bigl(\mathbf{H}^{(k)}\bigr)^{\top}
\mathbf{R}^{-1}\mathbf{H}^{(k)}
+\mathbf{W}_{p},
\\
\mathbf{b}^{(k)}
&=
\bigl(\mathbf{H}^{(k)}\bigr)^{\top}
\mathbf{R}^{-1}\mathbf{r}_{m}^{(k)}
+\mathbf{W}_{p}\mathbf{r}_{p}^{(k)},
\\
\mathbf{W}_{p}
&=
\boldsymbol{\Sigma}_{\mathrm{FGNN}}^{-1},
\qquad
\mathbf{r}_{m}^{(k)}
=
\mathbf{z}-\mathbf{h}\!\left(\mathbf{x}^{(k)}\right),
\qquad
\mathbf{r}_{p}^{(k)}
=
\boldsymbol{\mu}_{\mathrm{FGNN}}-\mathbf{x}^{(k)}.
\end{aligned}
\]
Here, $\mathbf{H}^{(k)}$ and $\mathbf{C}^{(k)}$ denote the measurement and
constraint Jacobian, respectively, while
$\boldsymbol{\lambda}^{(k)}$ represent the Lagrange multipliers associated
with the equality constraints. The state estimate is then updated as:
\begin{equation}
\mathbf{x}^{(k+1)}
=
\mathbf{x}^{(k)}
+
\alpha_k\Delta\mathbf{x}^{(k)},
\label{eq:smap_state_update}
\end{equation}

The FGNN prior regularises the constrained MAP objective and improves the numerical conditioning of the linear system solved at each iteration, while the step length $\alpha_k \in (0,1]$, selected through a line-search procedure, scales the Gauss--Newton correction to promote stable convergence and avoid excessively large update steps. Each refinement step moves the current estimate closer to the constrained MAP solution. After $K$ iterations, $\mathcal{S}_{\mathrm{MAP}}$ layer returns the final estimate $\mathbf{x}^{\star}_{\mathrm{MAP}}$. Rather than requiring full convergence of the inner solver, end-to-end training encourages the FGNN to produce an initialization and prior that can be refined into a physically constrained estimate within a limited number of fixed iterations. The final state estimate defines the linearization point for approximating the local posterior covariance, which characterizes the uncertainty around the final MAP estimate. The overall \(\mathcal{S}_{\mathrm{MAP}}\) estimation procedure is shown in Fig.~\ref{fig:s_map_flow}.

The formulation in \eqref{eq:gaussian_constrained_map} extends to non-Gaussian measurement uncertainty by replacing the quadratic measurement-residual term with the negative log-likelihood corresponding to the assumed measurement noise distribution, while retaining the same equality constraints. The resulting constrained MAP problem is solved using an iteratively reweighted least-squares (IRLS) procedure~\cite{zhao2017gmukf}. At each iteration, the measurement weights are updated based on the current residual values, and the resulting equality-constrained weighted least-squares subproblem is solved using the same KKT formulation. Importantly, the IRLS procedure does not require explicit specification of the true measurement noise distribution and can accommodate a broad range of non-Gaussian and heteroscedastic measurement uncertainties through iterative reweighting. The overall framework exploits sparse linear algebra and supports batched GPU execution in both the forward and backward passes. Additionally, the $\mathcal{S}_{\mathrm{MAP}}$ layer differs from conventional projection-based approaches by treating the neural network output as a learned prior within an end-to-end differentiable constrained MAP formulation, thereby jointly balancing prior information, measurement likelihood, and physical constraints during refinement.


\subsection{Geometric Surrogate-Gradient Method}
\label{surrogate_method}

End-to-end model training requires propagating the gradient of the training loss through the $\mathcal{S}_{\mathrm{MAP}}$ iterations to the FGNN. This can be achieved either by explicitly unrolling the solver iterations or by applying implicit differentiation to the local optimality conditions. Explicit unrolling is computationally and memory intensive and may yield unstable gradients~\cite{metz2019understanding}, whereas implicit differentiation requires solving an additional linear system involving the Jacobian of the fixed-point or optimality conditions at the solution, and generally assumes that the forward solve has sufficiently converged and that the resulting linear system is well conditioned~\cite{blondel2022efficient}. These requirements can be difficult to satisfy in nonlinear DSSE under a fixed solver iteration, sparse measurements, and poor observability, potentially leading to unstable or inaccurate gradient computations. 

\begin{figure}[!t]
  \centering
  \includegraphics[width=0.75\linewidth,trim=10pt 3pt 10pt 10pt,clip]{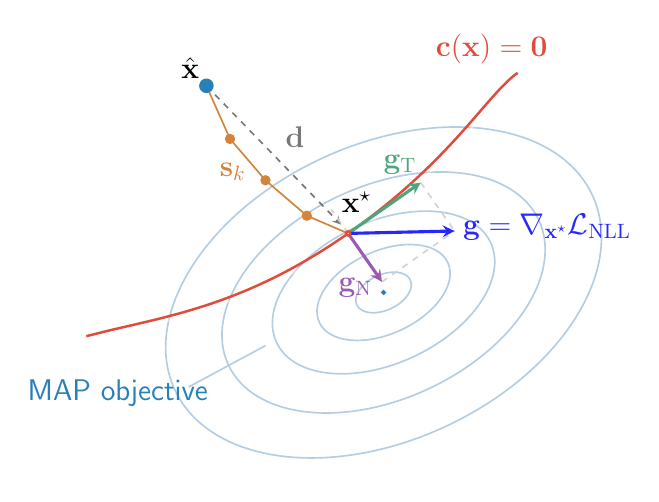}
    \caption{Geometric interpretation of the surrogate backward pass. From the FGNN prediction
    $\hat{\mathbf{x}}$, the constrained $\mathcal{S}_{\mathrm{MAP}}$ layer follows steps $\mathbf{s}_{k}$ to
    the feasible estimate $\mathbf{x}^{\star}$ on
    $\mathbf{c}(\mathbf{x})=\mathbf{0}$, with total displacement $\mathbf{d}$. The loss gradient with respect to the refined state,
    $\mathbf{g}=\nabla_{\mathbf{x}^{\star}}\mathcal{L}_{\mathrm{NLL}}$, is decomposed into
    normal and tangential components. The normal component, which would move the
    estimate off the feasible set toward the unconstrained optimum, is discarded,
    while the tangential component is backpropagated to the FGNN.}
  \label{fig:geom_surrogate}
\end{figure}

To address this, HCP-PINN adopts a rank-$r$ geometric surrogate-gradient method that propagates a stable, geometry-aware gradient signal from $\mathcal{S}_{\mathrm{MAP}}$ to the FGNN, guiding it to to produce state estimates that are more physically consistent. Starting from the FGNN prediction $\hat{\mathbf{x}}$, the constrained
$\mathcal{S}_{\mathrm{MAP}}$ layer returns the refined estimate
$\mathbf{x}^{\star}$, with displacement
$\mathbf{d}=\mathbf{x}^{\star}-\hat{\mathbf{x}}$. End-to-end differentiation through this refinement would ordinarily require unrolling the full sequence of solver updates $\mathbf{s}_k$ and retaining the corresponding computational graph during backpropagation. To avoid this overhead, HCP-PINN employs a rank-$r$ geometric surrogate-gradient method that constructs the backward signal from the local constraint geometry at the final estimate $\mathbf{x}^{\star}$, as illustrated in Fig.~\ref{fig:geom_surrogate}. 
At the refined estimate $\mathbf{x}^{\star}$, the constraint Jacobian
\begin{equation}
\mathbf{C}^{\star}
=\left.\frac{\partial\mathbf{c}(\mathbf{x})}{\partial\mathbf{x}}\right|_{\mathbf{x}=\mathbf{x}^{\star}}
\in\mathbb{R}^{c\times n},
\qquad r=\operatorname{rank}(\mathbf{C}^{\star})\leq \textbf{c},
\label{eq:final_constraint_jacobian}
\end{equation}
determines the local geometry of the linearized constraint manifold. Its row space defines the local constraint-normal directions, whereas its null space corresponds to the tangent directions along which the constraints remain satisfied to first order. Let $\mathbf{D}^{\star}\in\mathbb{R}^{n\times r}$ denote the orthonormal basis of the local constraint-normal space obtained from a rank-revealing QR factorization of $(\mathbf{C}^{\star})^{\top}$:
\begin{equation}
(\mathbf{D}^{\star})^{\top}\mathbf{D}^{\star}=\mathbf{I}_{r},
\qquad
\operatorname{range}(\mathbf{D}^{\star})
=
\operatorname{range}\!\left((\mathbf{C}^{\star})^{\top}\right).
\label{eq:normal_space_basis}
\end{equation}
where $\mathbf{I}_r$ denotes the $r\times r$ identity matrix. Writing $\mathbf{g}:=\nabla_{\mathbf{x}^{\star}}\mathcal{L}_{\mathrm{NLL}}$
for the training loss gradient evaluated at the final refined estimate, its normal
and tangential components are:
\begin{equation}
\mathbf{g}=\mathbf{g}_{\mathrm{N}}+\mathbf{g}_{\mathrm{T}},
\qquad
\mathbf{g}_{\mathrm{N}}=\mathbf{D}^{\star}(\mathbf{D}^{\star})^{\top}\mathbf{g},
\qquad
\mathbf{g}_{\mathrm{T}}
=\underbrace{\left(\mathbf{I}-\mathbf{D}^{\star}(\mathbf{D}^{\star})^{\top}\right)}_{\widetilde{\mathbf{J}}_{\hat{\mathbf{x}}}}\mathbf{g},
\label{eq:gradient_decomposition}
\end{equation}
where $\widetilde{\mathbf{J}}_{\mathbf{x}}$ is the orthogonal projector onto the local tangent space. It removes the constraint-normal component of the gradient while retaining the component tangent to the linearized constraint manifold. Accordingly, the surrogate gradient through $\mathcal{S}_{\mathrm{MAP}}$ is

\begin{equation}
\widetilde{\nabla}_{\hat{\mathbf{x}}}\mathcal{L}_{\mathrm{NLL}}
=
\widetilde{\mathbf{J}}_{\hat{\mathbf{x}}}\mathbf{g}
=
\mathbf{g}_{\mathrm{T}},
\qquad
\mathbf{C}^{\star}\mathbf{g}_{\mathrm{T}}=\mathbf{0},
\label{eq:surrogate_tangent_gradient}
\end{equation}
such that $\mathbf{g}_{\mathrm{T}}$ lies in the local tangent space and preserves the linearized equality constraints to first order. Conversely, the discarded normal component $\mathbf{g}{\mathrm{N}}$ represents the part of the training signal that would move the estimate away from the  constraint manifold. Finally, the gradient propagated to the FGNN combines the surrogate gradient of the NLL term with the direct gradient contribution from the auxiliary consistency loss:
\begin{equation}
\boxed{\;
\widetilde{\nabla}_{\hat{\mathbf{x}}}\mathcal{L}
=
\mathbf{g}_{\mathrm{T}}
+
\gamma
\frac{\partial\mathcal{L}_{\mathrm{cons}}}
{\partial\hat{\mathbf{x}}}
\;}
\label{eq:total_fgnn_gradient}
\end{equation}
Empirically, the proposed geometric-surrogate backward method provides a more efficient alternative to exact solver unrolling by reducing computational and memory costs during training, while achieving comparable estimation accuracy with fewer $\mathcal{S}_{\mathrm{MAP}}$ refinement iterations. The end-to-end training procedure of the proposed HCP-PINN is summarized in Algorithm~\ref{alg:hcppinn_training}.
%


\begin{algorithm}[!t]
\caption{HCP-PINN Training Algorithm}
\label{alg:hcppinn_training}
\begin{algorithmic}[1]
\State \textbf{Initialize:} FGNN parameters $\phi$ and estimation layer $\mathcal{S}_{\mathrm{MAP}}$ parameters
\Repeat
    \State Input a mini-batch of noisy, incomplete measurements $\mathbf{z}$
    \State Construct the factor graph and perform FGNN message passing
    \State Predict the state prior
    $\left(
    \boldsymbol{\mu}_{\mathrm{FGNN}},
    \boldsymbol{\Sigma}_{\mathrm{FGNN}}
    \right)
    =
    f_{\phi}(\mathbf{z},\mathcal{G}_f)$
    \State Compute the refined state estimate
    $\mathbf{x}^{\star}
    =
    \mathcal{S}_{\mathrm{MAP}}
    \left(
    \boldsymbol{\mu}_{\mathrm{FGNN}},
    \boldsymbol{\Sigma}_{\mathrm{FGNN}},
    \mathbf{z}
    \right)$
    \State Evaluate the training loss $\mathcal{L(\phi)}$ using the chosen likelihood objective
    \State Backpropagate through $\mathcal{S}_{\mathrm{MAP}}$ using the geometric surrogate-gradient rule
    \State Update FGNN parameters $\phi$ using the Adam optimizer
\Until{convergence}
\end{algorithmic}
\end{algorithm}

\section{Experimental Setup}
\label{sec4}
\begin{figure}[!t]
  \centering
  \begin{subfigure}[b]{0.50\columnwidth}
    \centering
    \includegraphics[width=\linewidth,trim=15 35 5 25,clip]{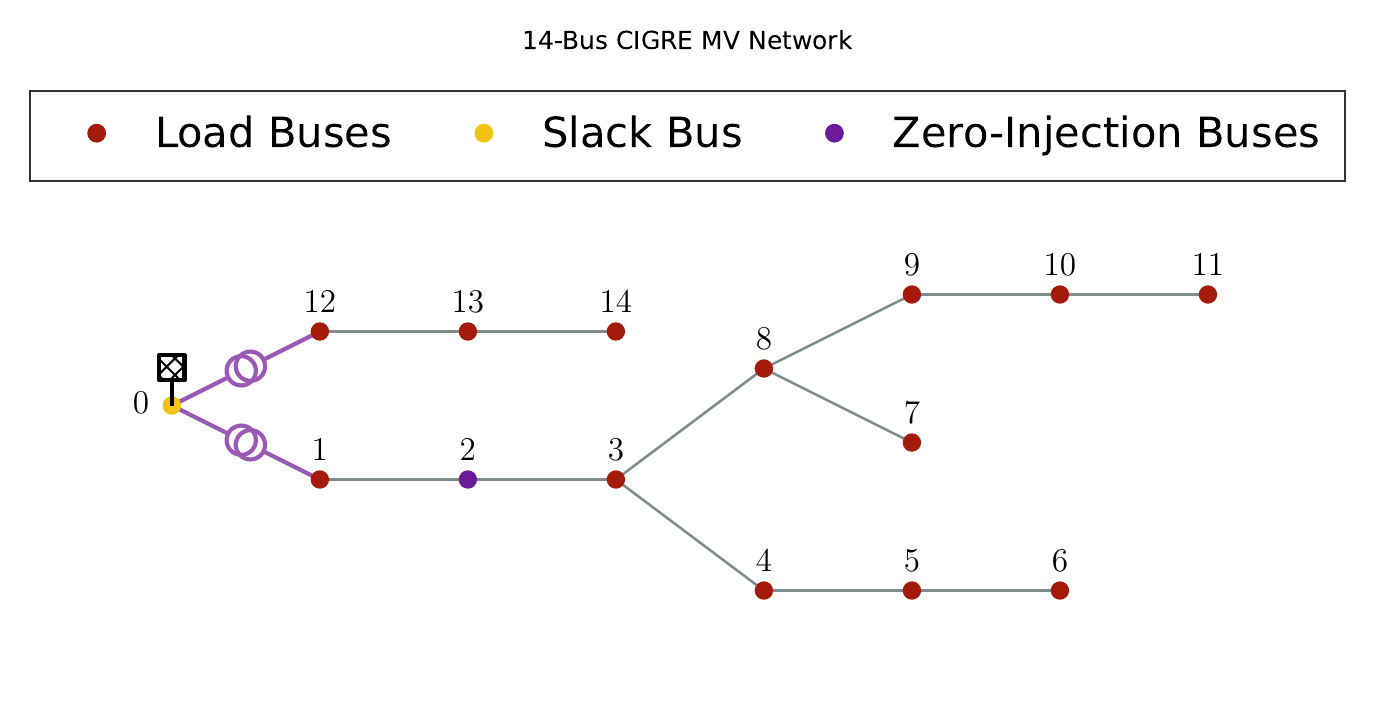}
    \vspace{-0.1in}
    \caption{14-Bus CIGRE MV Network}
    \label{fig:cigre_network}
  \end{subfigure}

  \begin{subfigure}[b]{\columnwidth}
    \centering
    \includegraphics[width=0.85\linewidth,trim=0 50 0 50,clip]{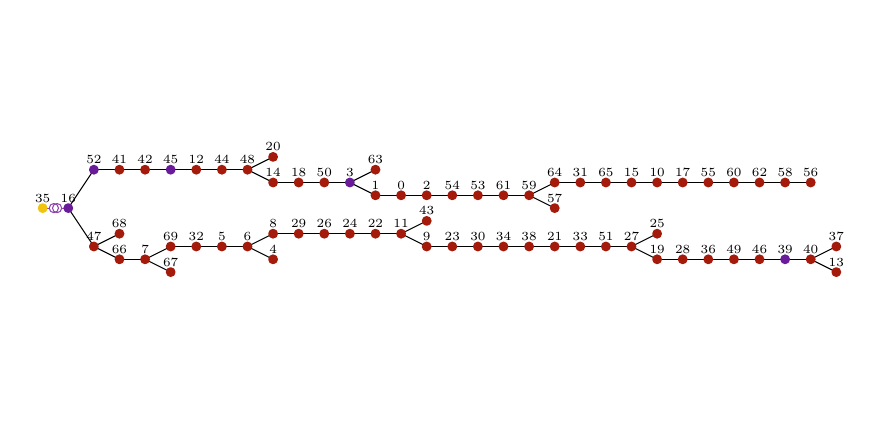}
    \vspace{-0.2in}
    \caption{70-Bus Oberrhein MV Network}
    \label{fig:oberrhein_network}
    \vspace{0.2in}
  \end{subfigure}

  \begin{subfigure}[b]{\columnwidth}
    \centering
    \includegraphics[width=0.65\linewidth,trim=25 35 0 32,clip]{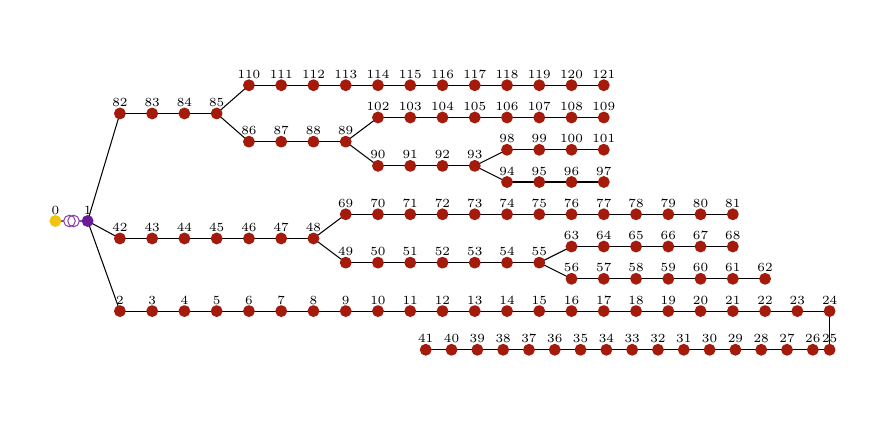}
    \caption{122-Bus Dickert LV Network}
    \label{fig:net3_network}
  \end{subfigure}
    \caption{Balanced test networks used in this study: (a)~14-bus CIGRE MV, (b)~70-bus Oberrhein MV, and (c)~122-bus Dickert LV.
      Buses are colored by type: load ({\color[HTML]{9E1A1A}$\bullet$}),
      slack ({\color[HTML]{F1C40F}$\bullet$}),
      and zero-injection ({\color[HTML]{702082}$\bullet$}).}
  \label{fig:test_networks}
\end{figure}

\begin{figure}[!t]
    \centering
    \begin{subfigure}{0.49\linewidth}
        \centering
        \includegraphics[width=\linewidth, trim={45 55 0 0}, clip]{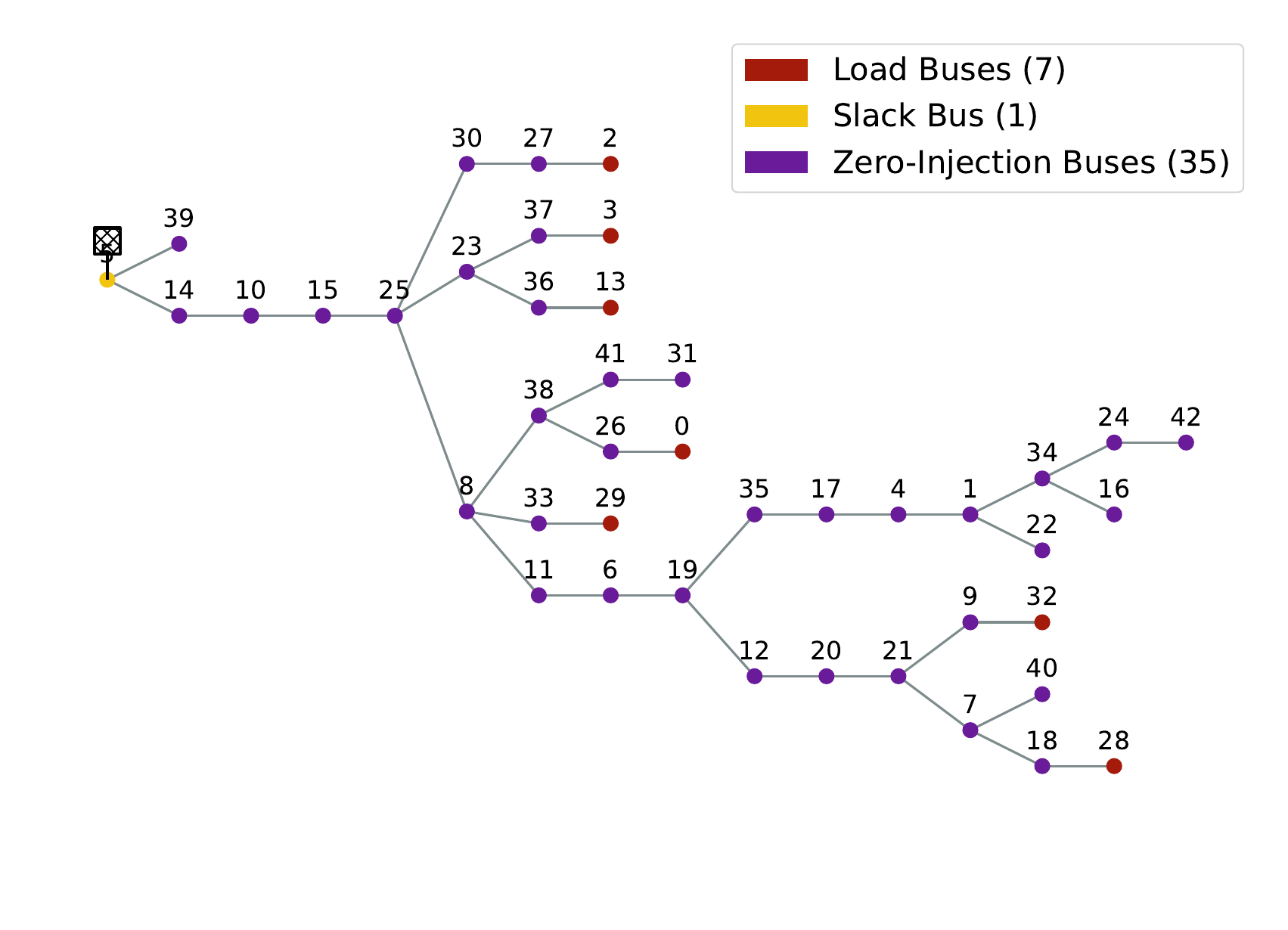}
        \caption{LV2-43bus feeder}
        \label{fig:network1}
    \end{subfigure}
    \hfill
    \begin{subfigure}{0.5\linewidth}
        \centering
        \includegraphics[width=\linewidth, trim={45 40 20 0}, clip]{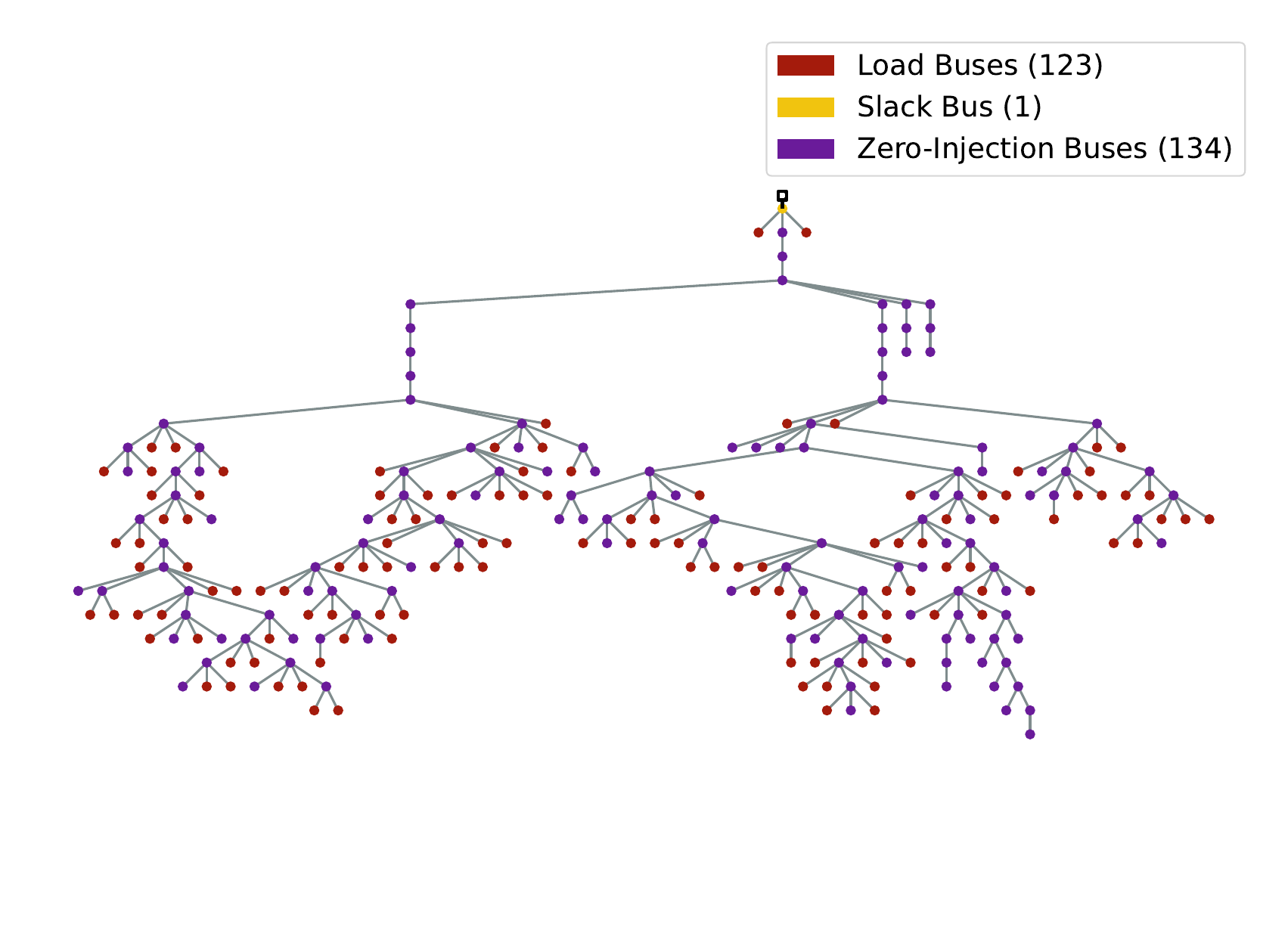}
        \caption{LV9-258bus feeder}
        \label{fig:network2}
    \end{subfigure}
    \caption{Three-phase unbalanced LV feeders: (a) LV2 with 43 buses and (b) LV9 with 258 buses.}
    \label{fig:3p_test_feeders}
\end{figure}

To demonstrate the effectiveness of the proposed HCP-PINN approach, a series of case studies is conducted under multiple forms of non-Gaussian measurement noise on balanced MV and unbalanced three-phase LV networks. The MV balanced test networks, selected from~\cite{thurner2018pandapower}, vary in topology, size, and PV penetration. They are included to facilitate direct comparison with conventional single phase-equivalent DSSE implementations available in widely used power system analysis tools such as \texttt{pandapower}. To further evaluate the accuracy and robustness of the proposed model under more realistic unbalanced operating conditions, two three-phase unbalanced LV feeders from a realistic Australian distribution grid, comprising 43 and 258 buses, respectively, are also considered~\cite{geth2025realisticAusMV}. Table~\ref{tab:test_systems} summarizes their key characteristics, while Figs.~\ref{fig:test_networks} and~\ref{fig:3p_test_feeders} shows their topological layouts. These test networks are representative of typical MV and LV distribution systems commonly used in DSSE studies.

The performance of the HCP-PINN is compared with numerical WLS, LAV, soft-constrained PINN, and purely data-driven GNN methods. Model performance is examined under varying levels of measurement noise (i.e., low, normal, and high) and different degrees of network observability (i.e., minimal, partial, and full), as detailed in Tables~\ref{tab:observability_levels} and~\ref{tab:noise_levels}. Following standard DSSE practice, the network topology and line parameters are assumed to be known during model training.
\begin{table}[!t]
  \centering
  \footnotesize
  \setlength{\tabcolsep}{3.0pt}
  \renewcommand{\arraystretch}{1.1}
  \caption{Characteristics of the test networks used in this study. OH and UG denote overhead and underground feeders, respectively.}  \label{tab:test_systems}
  \begin{tabular}{lccccccc}
    \toprule
    \textbf{Network} & \textbf{Type} & \textbf{Feeder} & \textbf{Base (kV)} & \textbf{\# Buses} & \textbf{\# Lines} & \textbf{\# Loads} & \textbf{\# PVs} \\
    \midrule
    14-bus      & Balanced & UG & 20   & 14  & 13  & 18  & 9  \\
    70-bus      & Balanced & UG & 20   & 70  & 69  & 61  & 60 \\
    122-bus     & Balanced & UG & 0.4  & 122 & 120 & 120 & 60 \\
    LV2-43bus   & 3$\phi$-unbalanced & OH    & 0.4 & 43  & 42  & 7   & 4 \\
    LV9-258bus  & 3$\phi$-unbalanced & OH & 0.4 & 258 & 257 & 123 & 62 \\
    \bottomrule
  \end{tabular}
\end{table}
Observability levels are defined in terms of \emph{fraction of available
measurements} (FAM) value. Let $\mathcal{M}_{\mathrm{full}}$ denote the complete set of candidate real measurements of a given type, enumerated over all corresponding buses,
phases, and branches, and let
$\mathcal{M}_{\mathrm{avail}} \subseteq \mathcal{M}_{\mathrm{full}}$ denote the
subset actually available in a given operating sample or scenario. The FAM is defined as the cardinality ratio:
\begin{equation}
\mathrm{FAM}
=
\frac{\lvert\mathcal{M}_{\mathrm{avail}}\rvert}
     {\lvert\mathcal{M}_{\mathrm{full}}\rvert}
\in[0,1],
\label{eq:fam}
\end{equation}
where $\mathrm{FAM}=1$ corresponds to full availability of real measurements, while lower values represent reduced measurement coverage. The FAM ratio is computed using real measurements only. If these measurements are insufficient to satisfy the minimum observability criterion, active- and reactive-power injection pseudo-measurements $\mathcal{P}$ are added at unmonitored buses.
\begin{table}[!t]
  \centering
  \caption{Fraction of available measurements (FAM) by measurement type under each observability scenario.
  }
  \label{tab:observability_levels}
  \small
  \setlength{\tabcolsep}{6pt}
  \renewcommand{\arraystretch}{1.15}
  \begin{tabular}{lccccc}
    \toprule
    \textbf{Scenario} & \( \mathcal{M}_V \) & \( \mathcal{M}_\theta \) & \( \mathcal{M}_{P_i} \) & \( \mathcal{M}_{Q_i} \) & \( \mathcal{M}_{P_{{ij}}} \) \\
    \midrule
    Minimal  & 0.25 & 0.00 & 0.25 & 0.25 & 0.00 \\
    Partial  & 0.50 & 0.00 & 0.50 & 0.50 & 0.00 \\
    Full     & 1.00 & 0.00 & 1.00 & 1.00 & 0.00 \\
    \bottomrule
  \end{tabular}
\end{table}
\begin{table}[t]
  \centering
  \caption{Relative measurement-noise levels for the considered scenarios, expressed as percentages of the corresponding true measurement values \(z_{\mathrm{true}}\). Pseudo-measurements \(\mathcal{P}\) are assigned higher uncertainty than real measurements.}
  \label{tab:noise_levels}
  \begin{tabular}{lcccc}
    \toprule
    \textbf{Scenario}
    & \(\eta_{\mathcal{}_V}\) (\%)
    & \(\eta_{\mathcal{}_{P_i}}\) (\%)
    & \(\eta_{\mathcal{}_{Q_i}}\) (\%)
    & \(\eta_{\mathcal{P}}\) (\%) \\
    \midrule
    Low    & 0.5 & 1.0 & 1.0 & 5  \\
    Normal & 1.0 & 2.0 & 2.0 & 10 \\
    High   & 3.0 & 5.0 & 5.0 & 15 \\
    \bottomrule
  \end{tabular}
\end{table}
\subsection{Dataset Generation}
\begin{figure}[!t]
  \centering
  \includegraphics[width=0.95\linewidth,trim=17pt 0pt 17pt 0pt,clip]{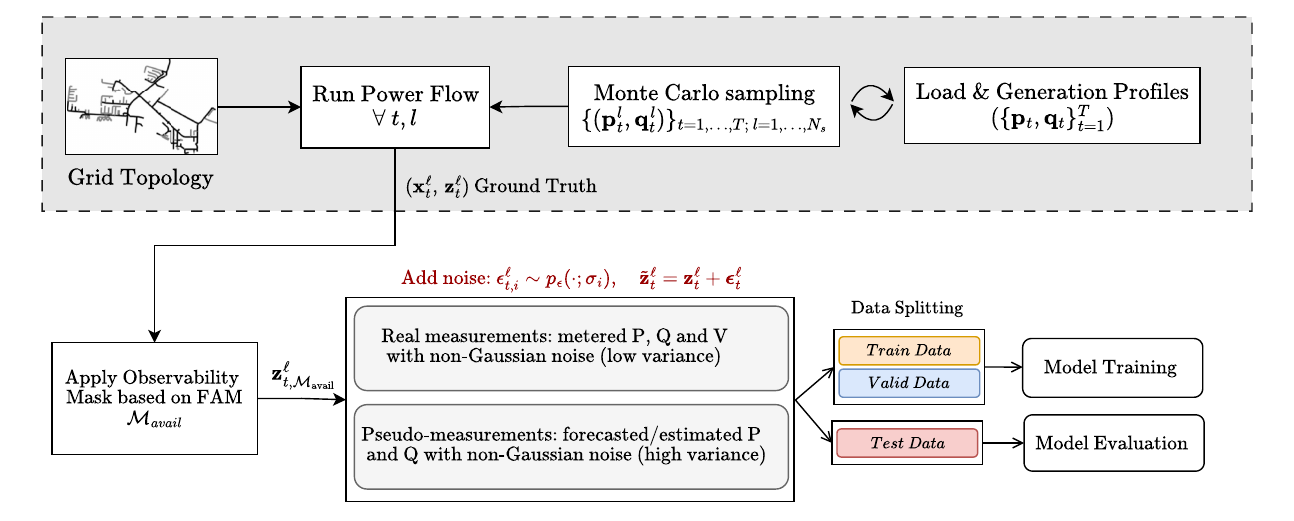}
  \caption{Dataset generation for model training and evaluation under noisy real and pseudo-measurements.}  
  \label{fig:data_flow}
\end{figure}

For each test network, $N_{\mathrm{days}}$ daily operating scenarios are generated from 24-hour load and generation profiles obtained from~\cite{r40}. To increase profile diversity, additive random perturbations with a standard deviation of $15\%$ are applied to the load profiles. Reactive power injections are then derived from the corresponding active power injections by assuming a constant power factor of $0.98$ for each residential customer. Each scenario $\ell\in\{1,\ldots,N_{\mathrm{days}}\}$ therefore comprises hourly active and reactive power injections $\{(\mathbf{p}_t^\ell,\mathbf{q}_t^\ell)\}_{t=1}^{24}$. The complete dataset generation pipeline is illustrated in Fig.~\ref{fig:data_flow}. 
To represent limited measurement availability, the FAM value determines the proportion of candidate measurements retained for each measurement type at a given observability level. The resulting measurement set is obtained by applying a binary mask $\mathbf{O}_t^\ell$, which gives:
\begin{equation}
\mathbf{z}_{t,\mathcal{M}_{\mathrm{avail}}}^{\ell}
=
\mathbf{O}_t^\ell\odot\mathbf{z}_t^\ell,
\qquad
\mathbf{O}_t^\ell\in\{0,1\}^{m},
\end{equation}
where $\odot$ denotes element-wise multiplication. The available real
measurements comprise voltage magnitudes $\mathcal{M}_{V}$ and active and
reactive power injections $\mathcal{M}_{P_i}$ and $\mathcal{M}_{Q_i}$.
Phase angle $\mathcal{M}_{\theta}$, active power flows $\mathcal{M}_{P_{{ij}}}$ and line current measurements are excluded, as they are generally not measured in LV networks. Subsequently, measurement noise is added according to the noise levels defined in Table~\ref{tab:noise_levels} and the noise distributions specified in Table~\ref{tab:noise_models}:
\begin{equation}
\tilde{z}_{t,i}^{\ell}
=
z_{t,i}^{\ell}
+
\varepsilon_{t,i}^{\ell},
\qquad
\varepsilon_{t,i}^{\ell}
\sim
p_{\varepsilon}(\cdot;\sigma_i).
\end{equation}
Noise levels for real measurements are defined to reflect typical smart-meter accuracy ranges, while pseudo-measurements are assigned larger, predominantly non-Gaussian errors to represent their lower reliability and greater forecasting uncertainty. This procedure generates the final dataset of noisy and incomplete measurement scenarios used to train and evaluate HCP-PINN under varying observability levels and non-Gaussian uncertainty. The dataset contains \(4320\) hourly operating scenarios and is partitioned into training, validation, and test sets using a \(60/20/20\) split.

\begin{figure}[!t]
  \centering
  \begin{subfigure}{0.32\columnwidth}
    \centering
    \includegraphics[width=\linewidth,trim=8 8 5 5,clip]{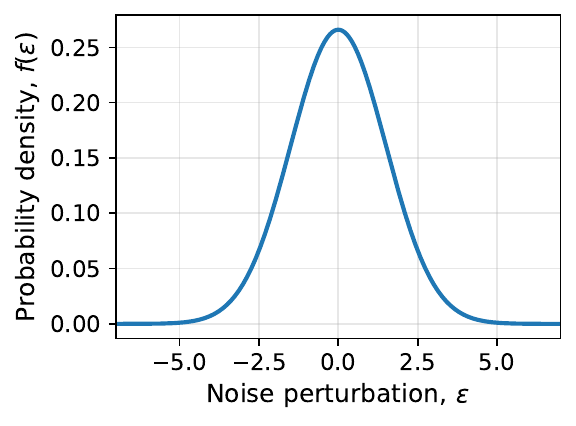}
    \caption{Gaussian}
    \label{fig:noise_gaussian}
  \end{subfigure}
  \hfill
  \begin{subfigure}{0.32\columnwidth}
    \centering
    \includegraphics[width=\linewidth,trim=8 8 5 5,clip]{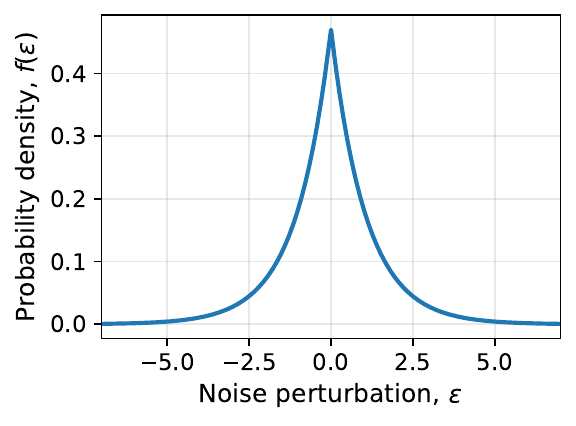}
    \caption{Laplace}
    \label{fig:noise_laplace}
  \end{subfigure}
  \hfill
  \begin{subfigure}{0.32\columnwidth}
    \centering
    \includegraphics[width=\linewidth,trim=8 8 5 5,clip]{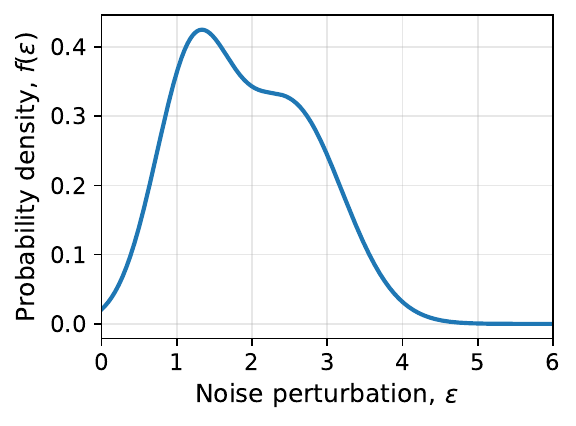}
    \caption{GMM}
    \label{fig:noise_gmm}
  \end{subfigure}
  \caption{Probability density functions of the considered measurement noise models.}
  \label{fig:noise_pdfs}
\end{figure}

\begin{table}[!t]
  \centering
    \caption{Measurement noise models, with the corresponding noise intensity levels reported in Table~\ref{tab:noise_levels}, where the standard deviation is defined as $\sigma=\dfrac{\eta}{100}|z_{\mathrm{true}}|$.}
  \label{tab:noise_models}

  \setlength{\tabcolsep}{8pt}
  \renewcommand{\arraystretch}{1.2}
  \small

  \begin{tabular}{@{} l c l @{}}
    \toprule
    \textbf{Noise Model} & \textbf{PDF} & \textbf{Parameters} \\
    \midrule

    Gaussian &
    $\displaystyle
    f(\varepsilon)=
    \frac{1}{\sqrt{2\pi}\sigma}
    \exp\!\left(-\frac{\varepsilon^2}{2\sigma^2}\right)$
    &
    $\mu=0,\quad
    \sigma$
    \\[12pt]

    Laplace &
    $\displaystyle
    f(\varepsilon)=
    \frac{1}{2b}
    \exp\!\left(-\frac{|\varepsilon|}{b}\right)$
    &
    $\displaystyle
    \mu=0,\quad
    b=\frac{\sigma}{\sqrt{2}}$
    \\[12pt]

    GMM &
    $\displaystyle
    f(\varepsilon)=
    \sum_{k=1}^{2}\pi_k
    \mathcal{N}(\varepsilon\mid\mu_k,\sigma_k^2)$
    &
    \begin{tabular}[c]{@{}l@{}}
      $\mu_1=1.2\sigma,\quad \mu_2=2.5\sigma$ \\[2pt]
      $\sigma_1=0.5\sigma,\quad \sigma_2=0.7\sigma$
    \end{tabular}
    \\

    \bottomrule
  \end{tabular}
\end{table}

\subsection{Non-Gaussian Uncertainty Models}
In practice, field measurement errors may deviate from conventional Gaussian assumptions because of meter accuracy differences, environmental effects, synchronization errors, communication disturbances, and bad data. Conventional numerical DSSE methods can achieve good accuracy when their modeling assumptions hold and the system is sufficiently observable. However, biased, heavy-tailed, or multimodal measurement noise, as well as outliers, can introduce model mismatch and degrade estimation performance. Accordingly, the proposed HCP-PINN is evaluated under representative Gaussian and non-Gaussian noise models. Measurements are corrupted by: (i) Gaussian noise as the baseline, (ii) Laplacian noise to represent heavy-tailed errors, and (iii) a two-component GMM to represent multimodal uncertainty. The corresponding distribution parameters are summarized in Table~\ref{tab:noise_models}, and their probability density functions (PDFs) are shown in Fig.~\ref{fig:noise_pdfs}.

%

%
%
%
%
\subsection{Training Configurations and Hyperparameters}
HCP-PINN is trained using a curriculum learning strategy~\cite{bengio2009curriculum}, in which training progresses from an unconstrained amortized estimation stage to a subsequent constrained optimization stage. In the first stage, the backbone FGNN is trained independently using the NLL objective, without the differentiable estimation layer $\mathcal{S}_{\mathrm{MAP}}$. This allows the FGNN to first learn a stable state estimate mapping from the noisy inputs before introducing the constrained refinement through $\mathcal{S}_{\mathrm{MAP}}$. In the second stage, $\mathcal{S}_{\mathrm{MAP}}$ is enabled and the complete model is trained end-to-end using the geometric surrogate-gradient backward method. This staged training procedure provides a stable initialization for the $\mathcal{S}_{\mathrm{MAP}}$ and mitigates unstable gradients during early training, thereby improving overall convergence. Model hyperparameter tuning was initially performed using a coarse search over broad parameter ranges, followed by Bayesian optimization with Optuna~\cite{akiba2019optuna} around the best-performing configurations identified on the validation set. The final hyperparameter values are reported in Table~\ref{tab:training_config}.

\begin{table}[!t]
\centering
\caption{Training configuration and model hyperparameters.}
\label{tab:training_config}
\begin{tabular}{ll}
\hline
\textbf{Hyperparameter} & \textbf{Setting} \\
\hline
Optimizer & Adam with AMSGrad \\
Epochs (Stage~1 / Stage~2) & 500 (250 / 250) \\
Batch size & 16 \\
Learning rate / Dropout & $10^{-3}$ / 0.1 \\
Activation / Hidden units & ReLU / 64 \\
FGNN message passing layers & $N=5$ \\
$\mathcal{S}_{\mathrm{MAP}}$ iterations & $K=3$ \\
Consistency loss weight & $\gamma=0.1$ \\
\hline
\end{tabular}
\end{table}

\subsection{Benchmark Methods}
\label{subsec:benchmark_methods}

The proposed HCP-PINN is evaluated against different numerical estimators and DL-based baselines. During evaluation, all methods are provided with the same noisy and incomplete measurement inputs and are compared under identical operating scenarios. The considered benchmark methods are:

\begin{itemize}
    \item \textbf{WLS estimator:} A conventional nonlinear least-squares state estimator that minimizes measurement residuals weighted by their variances. Despite the development of advanced SE alternatives, WLS remains a standard DSSE formulation in research and practice due to its well-established numerical behavior and modest modeling requirements. It is also used in industrial tools such as the Dutch DSO Alliander's open-source Power Grid Model~\cite{r46}, and is therefore selected as a primary numerical benchmark in this work.  

    \item \textbf{LAV estimator:} A robust estimator that minimizes the absolute measurement residuals. Its $\ell_1$ loss is less sensitive to outliers and heavy-tailed errors than WLS, making it a suitable benchmark for DSSE under non-Gaussian noise.

    \item \textbf{PINN:} A soft-constrained PINN comprising five graph layers with 128 hidden units per layer and ReLU activations. Its training objective combines a WLS-based measurement-fitting loss with a constraint penalty term:
    \begin{equation}
    \mathcal{L}_{\mathrm{PINN}}
    =
    \mathcal{L}_{\mathrm{WLS}}
    +
    \beta\,\mathcal{L}_{\mathrm{PB}},
    \end{equation}
    where $\mathcal{L}_{\mathrm{WLS}}$ penalizes the weighted measurement residuals, $\mathcal{L}_{\mathrm{PB}}$ penalizes violations of the power balance-based equality constraints, and $\beta$ controls the contribution of this penalty term. A default value of $\beta=0.5$ is used. This benchmark enables direct comparison between the soft constraint enforcement of a standard PINN and the hard-constrained formulation of the proposed HCP-PINN.

    
    \item \textbf{GNN:} A fully supervised GNN with five graph layers, 128 hidden units per layer, and ReLU activations, trained by minimizing the mean squared error (MSE) between the predicted and ground truth system states:
    \begin{equation}
    \mathcal{L}_{\mathrm{MSE}}
    =
    \frac{1}{n}
    \sum_{i=1}^{n}
    \left[
    \left(v_i^{\mathrm{pred}}-v_i^{\mathrm{true}}\right)^2
    +
    \left(\theta_i^{\mathrm{pred}}-\theta_i^{\mathrm{true}}\right)^2
    \right].
    \end{equation}
    
    The GNN is trained in a fully supervised manner using ground truth voltage magnitudes and phase angles obtained from AC power flow simulations. Since such complete state labels are generally unavailable in practical LV grids, the model is included as an optimistic supervised reference assuming access to complete ground truth state information during training.
    
    \end{itemize}


\subsection{Evaluation Metrics}
Model estimation accuracy is evaluated using the root mean square error
(RMSE):
\begin{equation}
\mathrm{RMSE}
=
\sqrt{
\frac{1}{N}
\sum_{i=1}^{N}
\left(
x_i^{\mathrm{PF}}
-
\hat{x}_i^{\mathrm{SE}}
\right)^2
},
\end{equation}
where \(N\) is the total number of evaluated state variables over all test
samples, \(x_i^{\mathrm{PF}}\) is the ground truth state obtained from the AC
power flow solution, and \(\hat{x}_i^{\mathrm{SE}}\) is the corresponding
estimated state. The accuracy and calibration of the HCP-PINN probabilistic state estimates are evaluated using the Continuous Ranked Probability Score (CRPS) metric:
\begin{equation}
  \mathrm{CRPS}(F,y)
  =
  \int_{-\infty}^{\infty}
  \left(F(u)-\mathbb{I}\{u\ge y\}\right)^2\,\mathrm{d}u,
\end{equation}
where \(F\) is the predicted cumulative distribution function, \(y\) is the ground truth value, \(u\) is the integration variable, and \(\mathbb{I}\{\cdot\}\) is the indicator function. CRPS is a proper scoring rule for evaluating continuous probabilistic predictions, with lower values indicating sharper and well-calibrated estimates.
\section{Numerical Results and Discussion}
\label{sec5}
The proposed HCP-PINN is evaluated across three balanced networks (14-bus, 70-bus, and 122-bus) and two unbalanced LV feeders (43-bus and 258-bus). The numerical evaluation assesses estimation accuracy under Gaussian and non-Gaussian measurement noise, constraint satisfaction, robustness to varying noise and observability levels, network model errors and topological reconfigurations, generalization to unseen loading conditions, uncertainty quantification, and computational efficiency.
\begin{figure}[!t]
  \centering
  \begin{subfigure}{0.48\linewidth}
    \centering
    \includegraphics[width=\linewidth]{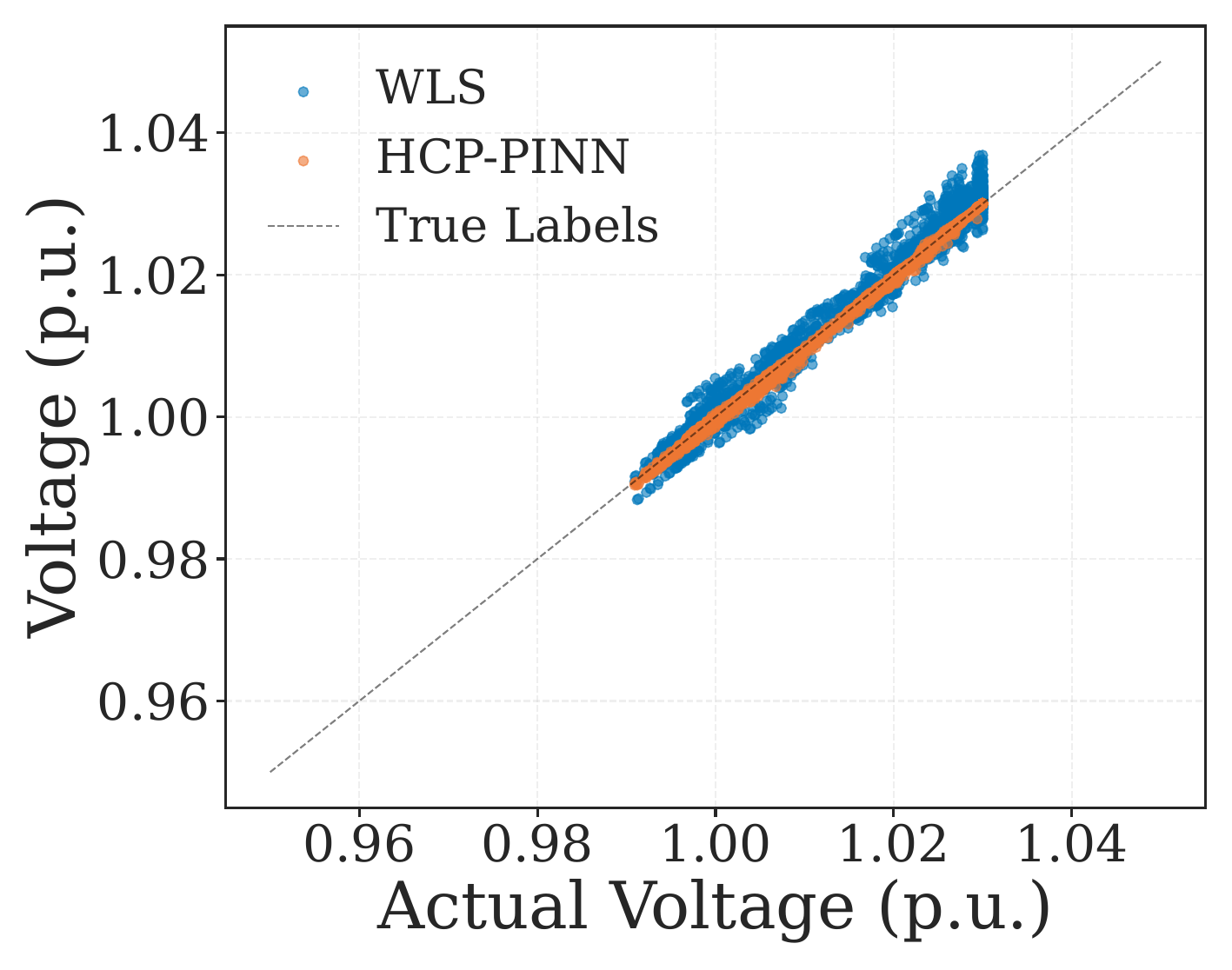}
    \caption{Voltage magnitude (p.u)}
  \end{subfigure}
  \hfill
  \begin{subfigure}{0.48\linewidth}
    \centering
    \includegraphics[width=\linewidth]{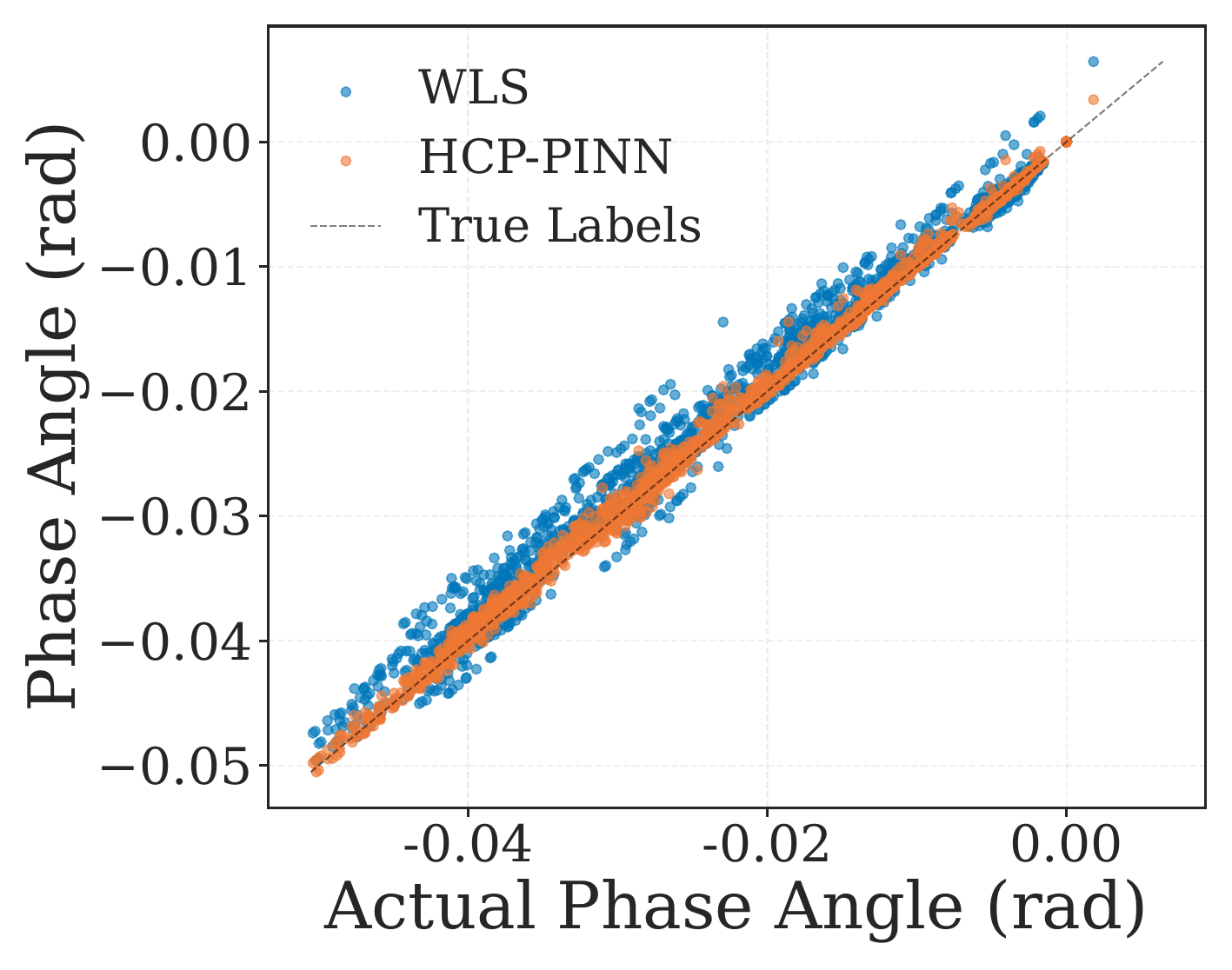}
    \caption{Phase angle (rad)}
  \end{subfigure}
  \vspace{0.3cm}

  \begin{subfigure}{0.48\linewidth}
    \centering
    \includegraphics[width=\linewidth]{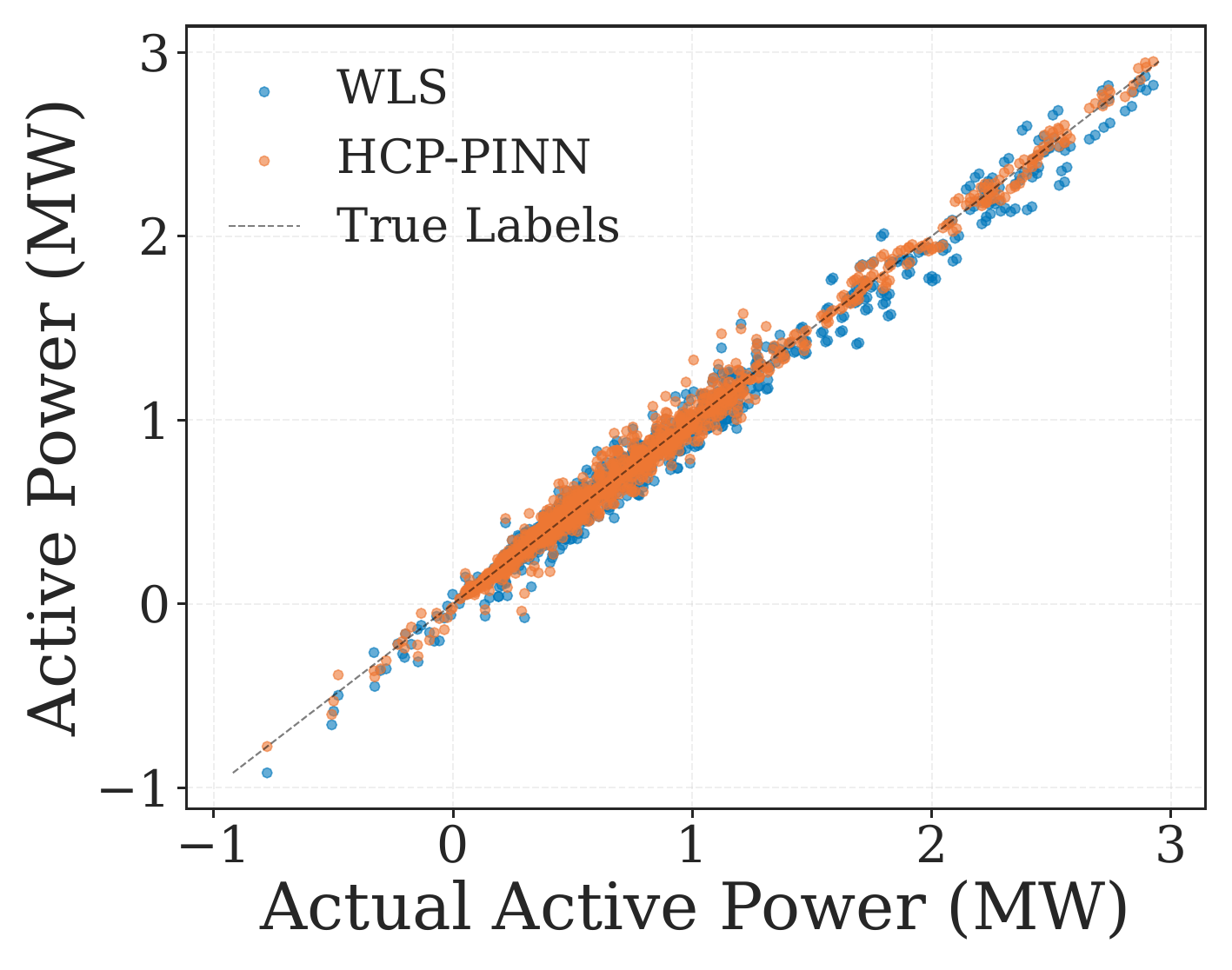}
    \caption{Active power flow (MW)}
  \end{subfigure}
  \hfill
  \begin{subfigure}{0.48\linewidth}
    \centering
    \includegraphics[width=\linewidth]{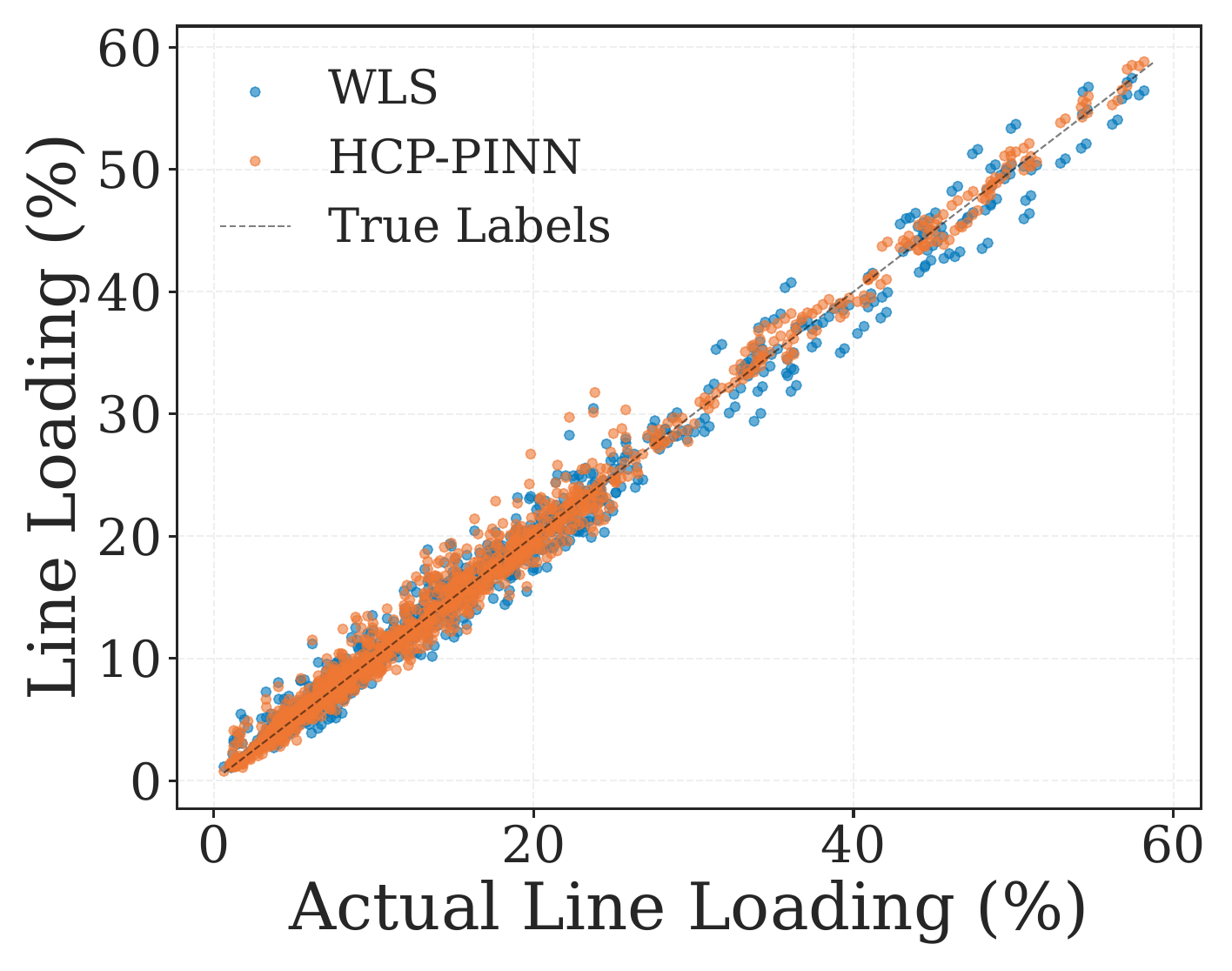}
    \caption{Line loading (\%)}
  \end{subfigure}

    \caption{Scatter plots of estimation errors under partial observability and Gaussian noise for the CIGRE 14-bus network}  \label{fig:scatter_comparison_2x2}
\end{figure}

\begin{table}[t]
\centering
\small
\setlength{\tabcolsep}{2.5pt}
\renewcommand{\arraystretch}{1.205}
\caption{Performance comparison averaged over all observability levels under Gaussian measurement noise.}
\label{tab:1ph_summary}

\begin{tabular}{l|l|ccccc}
\toprule
\textbf{Network} & \textbf{Quantity}
& \textbf{HCP-PINN} & \textbf{WLS} & \textbf{LAV}
& \textbf{PINN} & \textbf{GNN} \\
\midrule

\multirow{3}{*}{\textbf{14-bus}}
& $|V|$ RMSE ($\times 10^{-3}$)
& $\mathbf{0.69}$
& $2.03$ & $2.83$
& $2.9$ & $3.6$ \\

& $\theta$ RMSE ($\times 10^{-3}$)
& $\mathbf{0.48}$
& $1.33$ & $1.46$
& $2.4$ & $3.0$ \\

& Line loading (\%)
& $\textbf{0.90}$
& $0.95$ & $1.52$
& $5.8$ & $20.3$ \\

\midrule

\multirow{3}{*}{\textbf{70-bus}}
& $|V|$ RMSE ($\times 10^{-3}$)
& $\mathbf{0.41}$
& $2.02$ & $1.20$
& $4.1$ & $4.7$ \\

& $\theta$ RMSE ($\times 10^{-3}$)
& $\mathbf{0.27}$
& $0.92$ & $0.50$
& $5.2$ & $5.1$ \\

& Line loading (\%)
& $\mathbf{0.28}$
& $0.57$ & $0.32$
& $8.4$ & $28.4$ \\

\midrule

\multirow{3}{*}{\textbf{122-bus}}
& $|V|$ RMSE ($\times 10^{-3}$)
& $\mathbf{1.01}$
& $1.14$ & $1.53$
& $3.3$ & $3.5$ \\

& $\theta$ RMSE ($\times 10^{-3}$)
& $\mathbf{0.57}$
& $1.50$ & $1.44$
& $4.9$ & $2.5$ \\

& Line loading (\%)
& $1.62$
& $\mathbf{1.34}$ & $1.61$
& $15.6$ & $37.4$ \\

\bottomrule
\end{tabular}
\end{table}

\begin{table}[t]
\centering
\small
\caption{Per-phase RMSE for voltage magnitude ($|V|$, in $10^{-3}$\,p.u.),
phase angle ($\theta$, in $10^{-3}$\,rad), and line loading for the three-phase
LV2-43bus and LV9-258bus test networks.}
\label{tab:3ph_summary}
\renewcommand{\arraystretch}{1.15}
\setlength{\tabcolsep}{5pt}

\begin{tabular}{llrrrrrr}
\toprule
& & \multicolumn{3}{c}{\textbf{LV2-43 Feeder}}
  & \multicolumn{3}{c}{\textbf{LV9-258 Feeder}} \\
\cmidrule(lr){3-5}\cmidrule(lr){6-8}
\textbf{Model} & \textbf{Quantity} & A & B & C & A & B & C \\
\midrule

\multirow{3}{*}{\textbf{HCP-PINN}}
 & Voltage $|V|$ ($10^{-3}$)
 & \textbf{0.69} & \textbf{0.60} & \textbf{0.81}
 & \textbf{1.83} & \textbf{2.02} & \textbf{1.97} \\

 & Phase angle $\theta$ ($10^{-3}$)
 & \textbf{0.37} & \textbf{0.28} & \textbf{0.36}
 & 6.90 & 7.00 & 6.94 \\

 & Line loading (\%)
 & \textbf{0.40} & \textbf{0.51} & \textbf{0.47}
 & \textbf{1.23} & \textbf{1.45} & \textbf{1.27} \\

\cmidrule(lr){1-8}

\multirow{3}{*}{\textbf{WLS}}
 & Voltage $|V|$ ($10^{-3}$)
 & 7.91 & 7.32 & 7.42
 & 3.20 & 4.72 & 3.48 \\

 & Phase angle $\theta$ ($10^{-3}$)
 & 6.95 & 7.23 & 6.68
 & 14.80 & 17.90 & 37.50 \\

 & Line loading (\%)
 & 3.06 & 3.22 & 2.91
 & 1.76 & 1.91 & 1.82 \\

\cmidrule(lr){1-8}

\multirow{3}{*}{\textbf{LAV}}
 & Voltage $|V|$ ($10^{-3}$)
 & 6.67 & 6.81 & 8.43
 & 3.54 & 2.61 & 3.67 \\

 & Phase angle $\theta$ ($10^{-3}$)
 & 5.00 & 6.80 & 4.80
 & 9.36 & 10.90 & 9.54 \\

 & Line loading (\%)
 & 2.37 & 3.60 & 3.45
 & 3.05 & 3.41 & 3.04 \\

\cmidrule(lr){1-8}

\multirow{3}{*}{\textbf{PINN}}
 & Voltage $|V|$ ($10^{-3}$)
 & 1.40 & 0.87 & 1.86
 & 18.30 & 18.90 & 21.30 \\

 & Phase angle $\theta$ ($10^{-3}$)
 & 1.24 & 0.72 & 0.91
 & 5.61 & \textbf{5.59} & 5.59 \\

 & Line loading (\%)
 & 9.42 & 9.61 & 9.80
 & 22.30 & 22.80 & 23.20 \\

\cmidrule(lr){1-8}

\multirow{3}{*}{\textbf{GNN}}
 & Voltage $|V|$ ($10^{-3}$)
 & 1.40 & 1.12 & 1.50
 & 8.96 & 7.31 & 3.44 \\

 & Phase angle $\theta$ ($10^{-3}$)
 & 1.52 & 1.03 & 1.65
 & \textbf{5.59} & 5.69 & \textbf{4.73} \\

 & Line loading (\%)
 & 86.2 & 88.0 & 89.8
 & 168.0 & 171.0 & 175.0 \\

\bottomrule
\end{tabular}
\end{table}

\begin{figure}[!t]
  \centering
  \begin{subfigure}[t]{0.49\linewidth}
    \centering
    \includegraphics[
      width=\linewidth,
      trim=10pt 10pt 0pt 55pt,
      clip
    ]{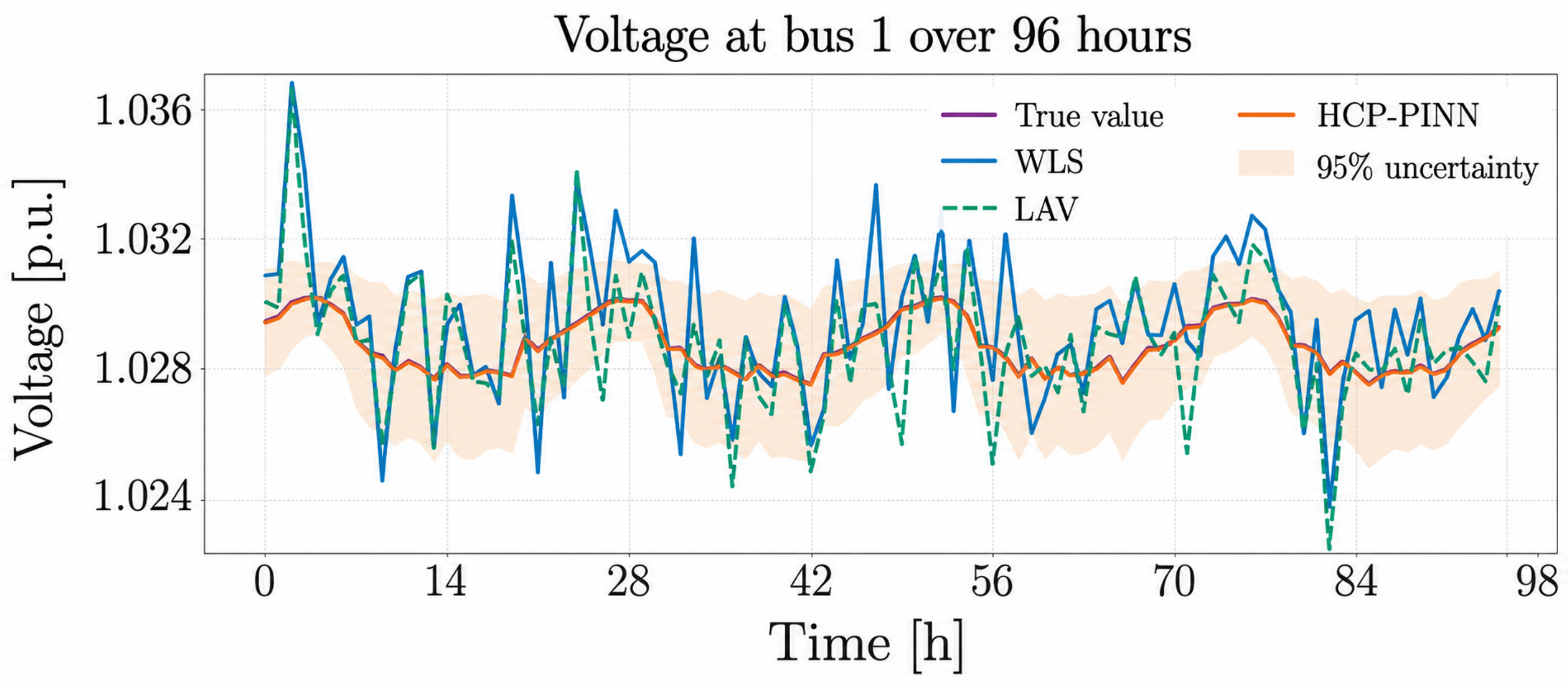}
    \caption{Bus~1 (unmonitored)}
    \label{fig:non_metered_bus}
  \end{subfigure}
  \hfill
  \begin{subfigure}[t]{0.49\linewidth}
    \centering
    \includegraphics[
      width=\linewidth,
      trim=10pt 10pt 0pt 30pt,
      clip
    ]{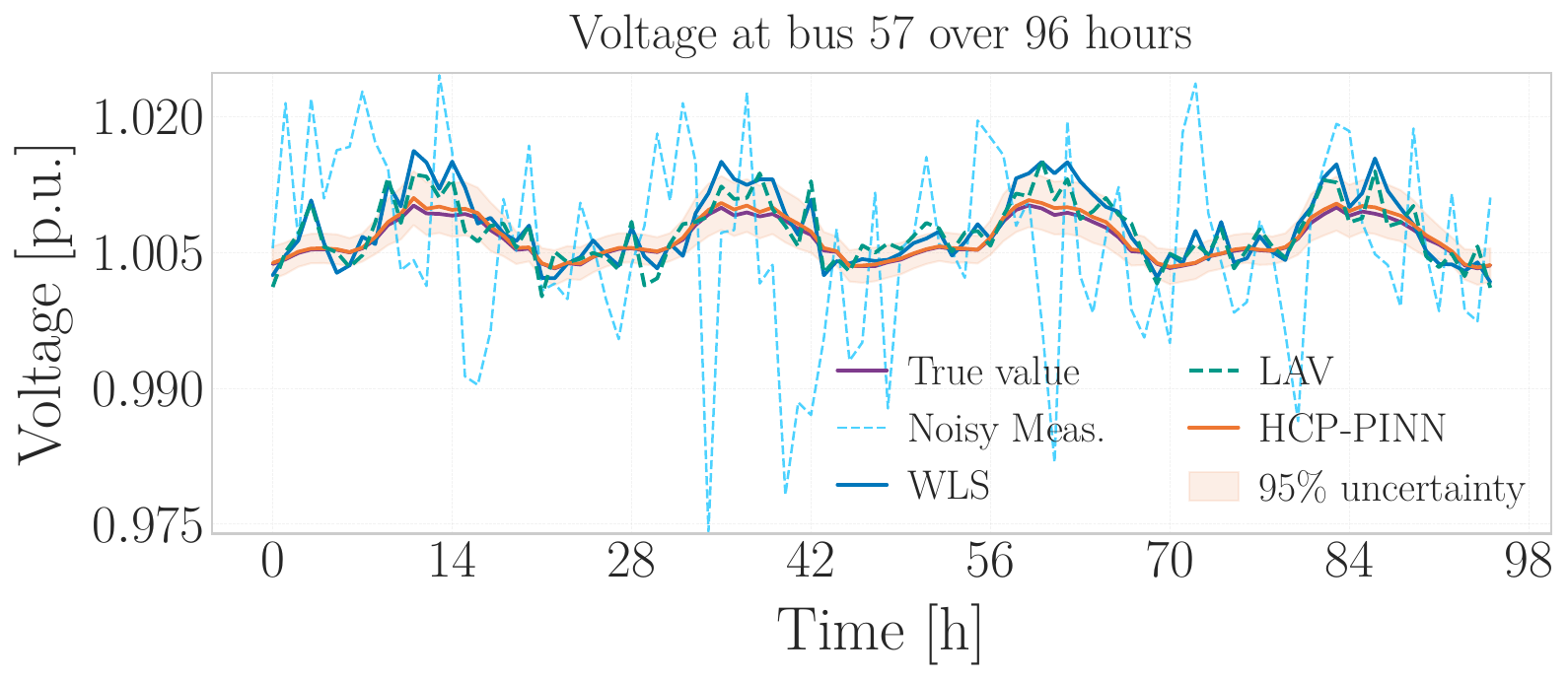}
    \caption{Bus~57 (monitored)}
    \label{fig:metered_bus}
  \end{subfigure}
  \label{fig:voltage_uncertainty_122bus}
    \caption{Voltage estimates at monitored and unmonitored buses in the
    70-bus network under high noise. HCP-PINN closely matches the
    ground truth voltages with calibrated 95\% confidence intervals.}
\label{fig:metered_bus_examples}
\end{figure}


%
\subsection{Estimation Accuracy under Gaussian Noise}
The estimation accuracy of the proposed HCP-PINN model is evaluated against all benchmark methods. Table~\ref{tab:1ph_summary} summarizes the RMSE for voltage magnitude, phase angle, and line loading across the MV balanced test networks. Overall, HCP-PINN consistently achieves lower estimation errors than WLS, LAV, PINN, and GNN benchmark methods. This is further illustrated in Fig.~\ref{fig:scatter_comparison_2x2}, where HCP-PINN predictions are more closely aligned with the ground truth $y=x$ diagonal line than those of WLS under partial observability on the CIGRE 14-bus network. The predictions also show a narrower spread and fewer large deviations across voltage magnitude, phase angle, line loading, and active power flows. Furthermore, Fig.~\ref{fig:metered_bus_examples} shows that HCP-PINN accurately recovers the true voltage magnitudes at both monitored and unmonitored buses in the 70-bus network, while demonstrating greater robustness to measurement noise than WLS and LAV.

Performance on unbalanced three-phase LV networks is evaluated using the LV2 43-bus and LV9 258-bus feeders. Table~\ref{tab:3ph_summary} reports the per-phase RMSE for all electrical quantities. HCP-PINN maintains lower estimation errors across both feeders, with particularly strong performance on the LV2-43 feeder. On the larger LV9-258 feeder, it outperforms WLS/LAV in voltage magnitude and line loading, while accuracy on phase angle estimates varies. The GNN achieves relatively low phase angle errors and reasonable voltage accuracy because it is trained directly on ground truth state labels. However, its extremely large line loading errors indicate that the model does not adequately learn the underlying physical relationships governing power flows and may therefore generalize poorly to unseen operating conditions. Similarly, the PINN model achieves relatively small phase angle errors but poor voltage magnitude and line loading estimates on the LV9-258 feeder, suggesting that penalty-based constraint enforcement may reduce power balance residuals without learning physically consistent state estimates. Line loading is directly used to assess congestion and overload risk, and bad estimates may therefore lead to incorrect operational decisions. In contrast, the hard-constrained estimation layer encourages the proposed HCP-PINN to learn physically consistent state estimates, which in turn improves the accuracy of derived quantities such as line loading.

\begin{figure}[!t]
  \centering
  \includegraphics[width=0.9\linewidth,trim=5pt 3pt 5pt 5pt,clip]{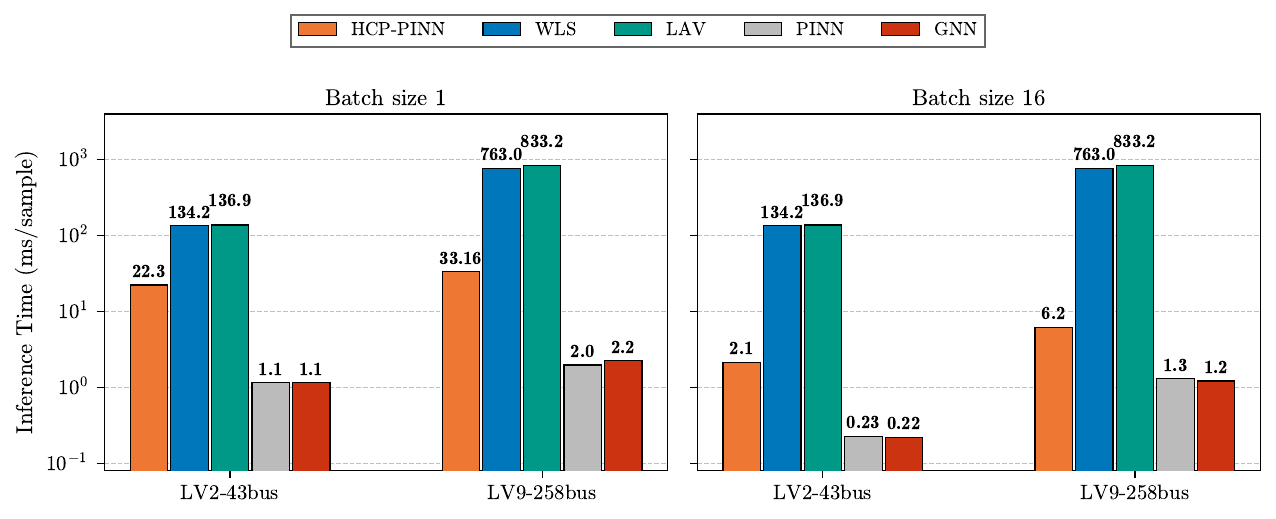}
    \caption{Per-sample computational time comparison for the LV2-43bus and LV9-258bus feeders under non-batched (batch size 1) and batched (batch size 16) execution.}
  \label{fig:runtime}
\end{figure}

Fig.~\ref{fig:runtime} compares the per-sample computational cost of all methods for both non-batched (batch size 1) and batched (batch size 16) evaluation. 
HCP-PINN is substantially faster than the numerical WLS and LAV estimators, with the computational advantage becoming more pronounced for the larger LV9-258bus feeder. In the non-batched case, HCP-PINN achieves speedups of up to approximately \(34\times\) over WLS and \(37\times\) over LAV on the LV9-258bus feeder. Batched execution further reduces the per-sample inference time of HCP-PINN from 22.3 ms to 2.1 ms for LV2-43bus and 6.2 ms for LV9-258bus, increasing the corresponding speedups on LV9-258bus to approximately \(123\times\) over WLS and \(134\times\) over LAV. This computational efficiency is enabled by the truncated iterative refinement, sparse linear algebra, and parallel GPU execution supported by the proposed HCP-PINN. The PINN and GNN baselines remain faster because they require only a direct forward pass without iterative refinement; however, this computational advantage comes at the cost of substantially lower estimation accuracy.

\begin{figure}[!t]
  \centering
  \begin{subfigure}{0.48\linewidth}
    \centering
    \includegraphics[width=\linewidth,trim=0pt 0pt 5pt 20pt,clip]{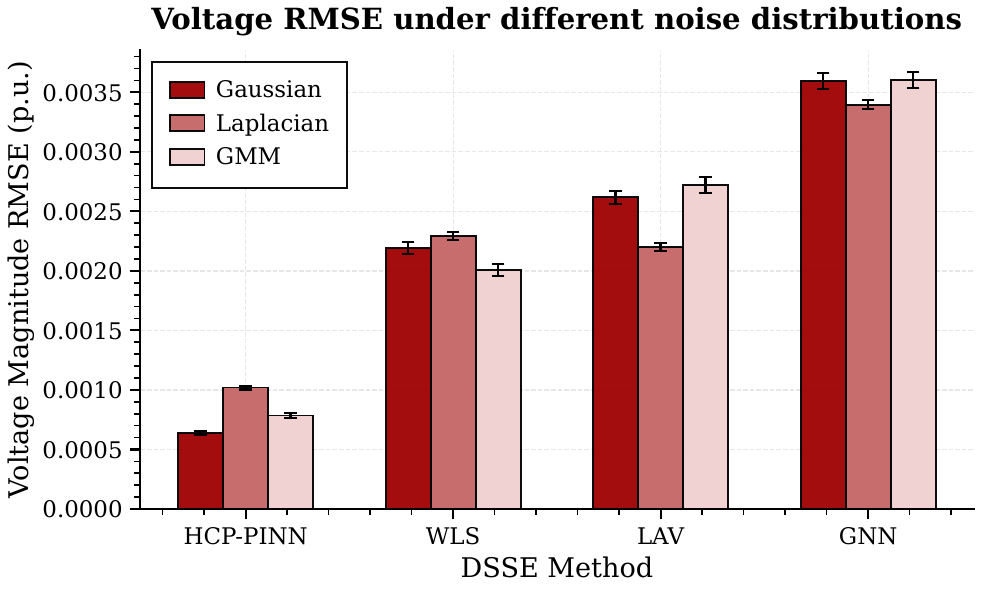}
    \caption{Voltage magnitude RMSE}
    \label{fig:noise_dist_vm}
  \end{subfigure}\hfill
  \begin{subfigure}{0.48\linewidth}
    \centering
    \includegraphics[width=\linewidth,trim=0pt 0pt 5pt 20pt,clip]{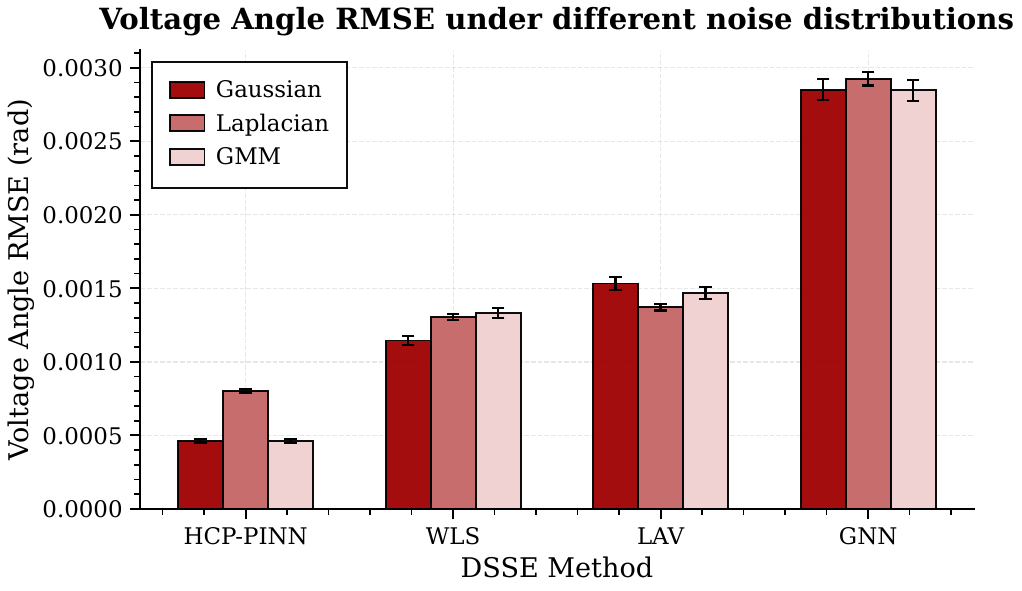}
    \caption{Phase angle RMSE}
    \label{fig:noise_dist_va}
  \end{subfigure}
  \caption{Voltage and phase angle RMSE under different measurement noise models for the 14-bus network with partial observability.}  \label{fig:noise_distribution_comparison_1x2}
\end{figure}

\subsection{Performance under Non-Gaussian Uncertainty}
In practice, the noise distribution of field measurements is often unknown and may deviate from the Gaussian assumption due to device errors, communication disturbances, and outlier contamination. Model robustness is therefore evaluated under three representative noise models on the CIGRE 14-bus test network: Gaussian noise as the baseline, Laplacian noise to represent heavy-tailed measurement errors, and a bimodal GMM to capture multimodal measurement uncertainty. As shown in Fig.~\ref{fig:noise_distribution_comparison_1x2}a, WLS shows a noticeable increase in voltage magnitude RMSE under heavy-tailed Laplacian noise, indicating greater sensitivity to non-Gaussian measurement errors. In contrast, the degradation under GMM noise is less pronounced. In contrast, LAV is more robust than WLS under heavy-tailed Laplacian noise due to its $\ell_1$ loss, although its performance still degrades under multimodal GMM noise. The proposed HCP-PINN, however, consistently achieves the lowest errors across all noise models, demonstrating stable performance under both heavy-tailed and multimodal measurement uncertainties. 

For phase angle estimation, shown in Fig.~\ref{fig:noise_distribution_comparison_1x2}b, no measurement noise is applied directly to the phase angles because they are not measured. Instead, the phase angles are inferred indirectly from the noisy $(|V|,P,Q)$ measurements. Their estimation errors therefore arise primarily from the propagation of noise in the available measurements rather than from direct angle measurement errors, resulting in different sensitivity patterns across noise models, particularly for WLS. The data-driven GNN exhibits substantially higher errors across all cases, indicating limited robustness and generalization under both Gaussian and non-Gaussian measurement noise.

\begin{figure}[!t]
  \centering
  \begin{subfigure}{0.48\linewidth}
    \centering
    \includegraphics[width=\linewidth,trim=0pt 0pt 0pt 11pt,clip]{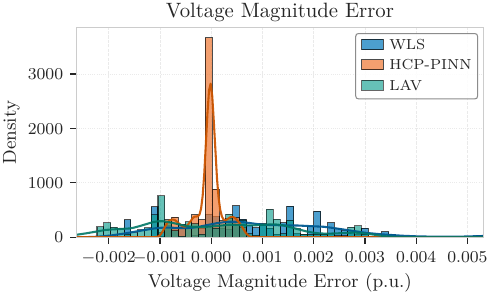}
    \caption{GMM noise}
    \label{fig:err_volt_gmm}
  \end{subfigure}\hfill
  \begin{subfigure}{0.48\linewidth}
    \centering
    \includegraphics[width=\linewidth,trim=0pt 0pt 0pt 11pt,clip]{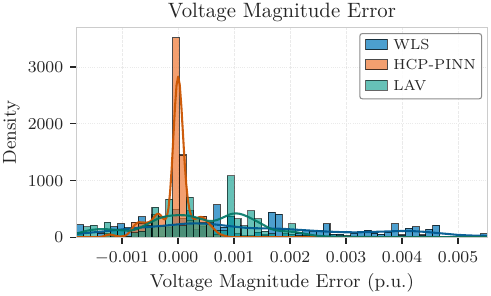}
    \caption{Laplace noise}
    \label{fig:err_volt_laplace}
    \end{subfigure}

  \caption{Estimated voltage error distribution comparison under GMM and Laplacian measurement noise for the 70-bus network.}
  \label{fig:error_distribution_voltage_1x2}
\end{figure}
%
%
%
Furthermore, robustness to non-Gaussian measurement errors is evaluated on the Oberrhein 70-bus network by examining the distribution of voltage estimation errors, with the same noise model applied to both real and pseudo-measurements in each case. Beyond the RMSE metric, this analysis characterizes changes in the full error distribution, including its concentration, spread, and tail behavior. As shown in Fig.~\ref{fig:error_distribution_voltage_1x2}, HCP-PINN errors remain tightly concentrated around zero under both GMM and Laplacian noise. In contrast, WLS exhibits more dispersed error distributions and a more pronounced positive tail under Laplacian noise, while LAV shifts from predominantly negative errors under GMM noise to positive errors under Laplacian noise. Although LAV is statistically matched to Laplacian measurement errors, this does not necessarily translate into lower estimation errors in nonlinear DSSE under limited observability. Overall, HCP-PINN maintains more consistent estimation performance across different forms of non-Gaussian uncertainty.
%


%
\subsection{Robustness to High Noise and Outliers}

\begin{figure}[!t]
  \centering
  \begin{subfigure}{0.48\linewidth}
    \centering
    \includegraphics[width=\linewidth,trim=0pt 0pt 0pt 0pt,clip]{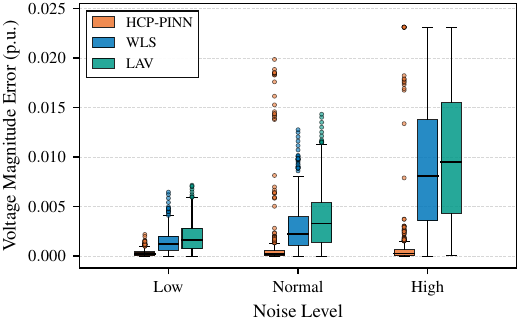}
    \caption{Voltage magnitude error under increasing Gaussian noise}
    \label{fig:noise_gaussian_vm}
  \end{subfigure}\hfill
  \begin{subfigure}{0.48\linewidth}
    \centering
    \includegraphics[width=\linewidth,trim=0pt 0pt 0pt 0pt,clip]{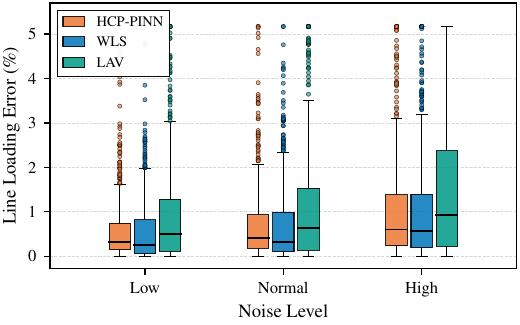}
    \caption{line loading error under increasing Gaussian noise}
    \label{fig:noise_gaussian_va}
  \end{subfigure}
  \caption{Impact of increasing measurement noise on the benchmark methods' accuracy under partial observability on the CIGRE 14-bus system.}
  \label{fig:noise_comparison_gaussian_1x2}
\end{figure}

Robustness to increasing measurement noise is evaluated under partial observability using three zero-mean Gaussian noise regimes---\textit{low}, \textit{normal}, and \textit{high}---defined in Table~\ref{tab:noise_levels}. Fig.~\ref{fig:noise_comparison_gaussian_1x2} shows the resulting voltage magnitude and line loading error distributions. As expected, these distributions generally show greater spread as noise intensity increases. However, HCP-PINN consistently achieves the lowest median errors and narrowest interquartile ranges compared with numerical WLS and LAV. The advantage is most pronounced under \textit{high} noise, where WLS and LAV show larger error spreads and more extreme outliers, while HCP-PINN maintains a comparatively narrower error distribution for both voltage magnitude and line loading. Overall, these results demonstrate the robustness of HCP-PINN under high measurement uncertainty and low signal-to-noise ratio conditions, while highlighting the greater sensitivity of conventional numerical estimators to measurement noise.

\begin{figure}[!t]

  \centering
  \begin{subfigure}{0.48\linewidth}
    \centering
    \includegraphics[width=\linewidth,trim=0pt 0pt 80pt 15pt,clip]{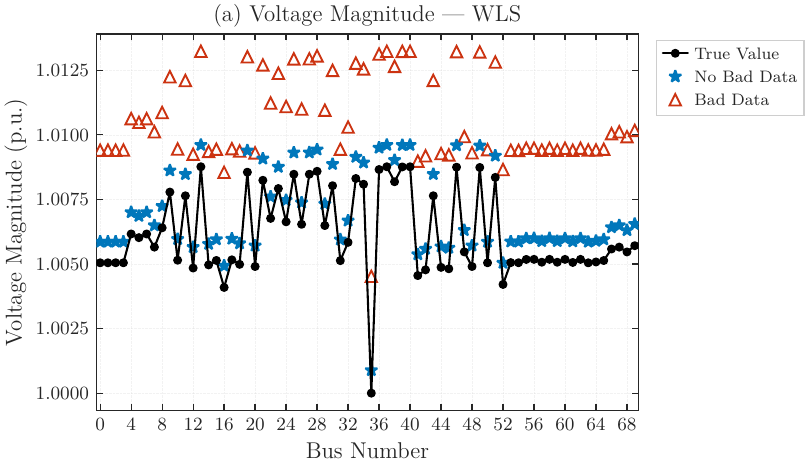}
    \caption{Numerical WLS}
    \label{fig:bad_data_gaussian_wls}
  \end{subfigure}\hfill
  \begin{subfigure}{0.48\linewidth}
    \centering
    \includegraphics[width=\linewidth,trim=0pt 0pt 80pt 15pt,clip]{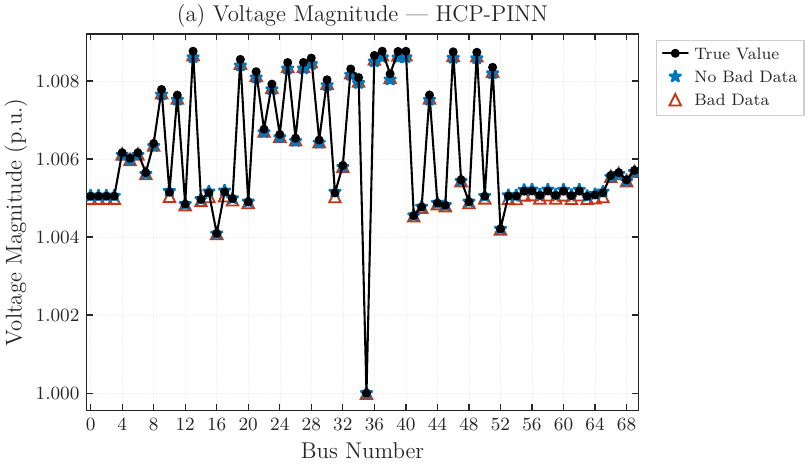}
    \caption{HCP-PINN}
    \label{fig:bad_data_gaussian_hcppinn}
  \end{subfigure}
  \caption{Bad data test under Gaussian noise for the 70-bus Oberrhein network: per-bus voltage magnitude obtained by numerical WLS and the proposed HCP-PINN method without ({\color{blue}\scriptsize$\star$}) bad data and with ({\color{red}\scriptsize$\triangle$}) bad data measurements, with ({\color{black}\tiny$\bullet$}) representing the ground truth values.}
  \label{fig:bad_data_exp}
\end{figure}

In addition, numerical WLS and HCP-PINN are compared under a controlled bad data scenario
in a partially observable setting with \(\mathrm{FAM}=0.5\). The active and
reactive power injection measurements at Bus~10 of the 70-bus network are
corrupted by scaling them to three times their true values. Fig.~\ref{fig:bad_data_gaussian_wls} compares the WLS voltage estimates obtained with ({\color{red}\scriptsize$\triangle$}) and without
({\color{blue}\scriptsize$\star$}) the corrupted measurements against the
ground truth ({\color{black}\tiny$\bullet$}). The corrupted injections substantially distort the WLS voltage profile because their assigned weights, based on the assumed measurement variances, do not account for the gross errors. Because WLS minimizes squared residuals, the large residuals introduced by a gross measurement error can dominate the objective, pulling the estimated state away from the true operating point and affecting voltage estimates at multiple buses.

In contrast, Fig.~\ref{fig:bad_data_gaussian_hcppinn} shows that HCP-PINN remains accurate across most buses in the considered bad data test scenario. This robustness can be attributed to the factor graph representation and physics-informed inductive biases of the model: the FGNN constructs a structured state prior by aggregating information across the factor graph, while the $\mathcal{S}_{\mathrm{MAP}}$ layer enforces the physical constraints during refinement steps. Together, these mechanisms restrict the solution toward physically consistent states and reduce the influence of outliers or corrupted measurements. These results provide two key insights. First, WLS is highly sensitive to bad data and typically requires a separate bad data detection and removal stage. Second, HCP-PINN is less affected in the considered evaluation scenario and remains accurate without an explicit bad data processing stage. While HCP-PINN shows strong empirical robustness in the considered bad data scenario, its performance under multiple simultaneous gross measurement errors remains to be evaluated.
\begin{figure}[!t]
  \centering
  \begin{subfigure}{0.48\linewidth}
    \centering
    \includegraphics[width=\linewidth,trim=0pt 0pt 0pt 0pt,clip]{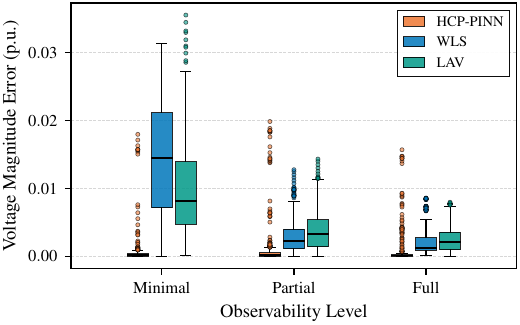}
    \caption{Voltage magnitude}
    \label{fig:obs_gaussian_voltage}
  \end{subfigure}\hfill
  \begin{subfigure}{0.48\linewidth}
    \centering
    \includegraphics[width=\linewidth,trim=0pt 0pt 0pt 0pt,clip]{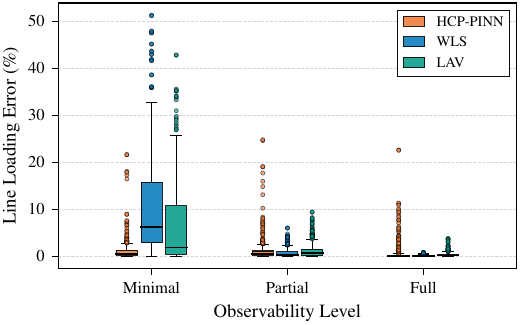}
    \caption{Line loading}
    \label{fig:obs_gaussian_loading}
  \end{subfigure}
  \caption{Impact of measurement coverage on model performance for the 70-bus network.}
  \label{fig:observability_comparison_gaussian_1x2}
\end{figure}
\subsection{Impact of Measurement Coverage}
The effect of measurement availability is assessed under three observability
regimes (i.e. \emph{minimal}, \emph{partial}, and \emph{full}) corresponding to
FAM values of \(0.25\), \(0.50\), and \(1.00\), respectively, for \(|V|\),
\(P_i\), and \(Q_i\), as summarized in Table~\ref{tab:observability_levels}. Lower FAM values indicate fewer real measurements and greater reliance on pseudo-measurements. In this experiment, errors in both real and pseudo-measurements are modeled as zero-mean Gaussian noise. HCP-PINN is
trained at \(\mathrm{FAM}=0.5\) and evaluated across all three regimes. Fig.~\ref{fig:observability_comparison_gaussian_1x2} presents the resulting
error distributions for (a) voltage magnitude and (b) line loading, comparing
WLS, LAV, and HCP-PINN. As measurement availability decreases, WLS and LAV show increasing median errors and greater dispersion due to the reduced availability of direct measurements and increased reliance on uncertain pseudo-measurements. This degradation is particularly evident under \emph{minimal} and \emph{partial} observability, where their boxplots exhibit wider interquartile ranges and more extreme errors. In contrast, HCP-PINN maintains comparatively small errors across the different observability levels, with its advantage over numerical estimators becoming clearer under the minimal observability scenario. This robustness can be attributed to the complementary inductive biases of the FGNN and the hard-constrained $\mathcal{S}_{\mathrm{MAP}}$ layer. The factor graph representation aligns with the underlying SE structure, enabling message passing to learn dependencies between measurements and system states and to propagate information across electrically coupled buses. This provides an informative state prior even when direct measurements are sparse. The $\mathcal{S}_{\mathrm{MAP}}$ layer then combines this prior with the available measurements while explicitly enforcing the equality constraints, which provide additional physical information and restrict the solution to feasible states. Consequently, HCP-PINN is less dependent on uncertain pseudo-measurements as measurement coverage decreases than WLS and LAV. This advantage is particularly evident for line loading under minimal observability, where HCP-PINN shows smaller overall errors than WLS and LAV.
\begin{table}[!t]
\centering
\small
\caption{Equality-constraint satisfaction at zero-injection buses under Gaussian noise and partial observability. Smaller power balance residuals, reported in kW, indicate better constraint satisfaction.}

\label{tab:zi_pb}
\renewcommand{\arraystretch}{1.25}
\setlength{\tabcolsep}{7pt}

\begin{tabular}{ccccc}
\toprule
\textbf{Network} & \textbf{Method} & \textbf{Mean}
& \textbf{95$^{\mathrm{th}}$ Percentile} & \textbf{Maximum} \\
\midrule

\multirow{4}{*}{\textbf{LV2 43-bus}}
& HCP-PINN
& \cellcolor{green!30} $\mathbf{2.51{\times}10^{-11}}$
& \cellcolor{green!30} $\mathbf{1.30{\times}10^{-9}}$
& \cellcolor{green!30} $\mathbf{1.83{\times}10^{-9}}$ \\[2pt]

& WLS
& \cellcolor{green!12} $9.17{\times}10^{-4}$
& \cellcolor{green!12} $5.92{\times}10^{-3}$
& \cellcolor{green!12} $8.45{\times}10^{-3}$ \\[2pt]

& PINN
& \cellcolor{orange!22} $3.94{\times}10^{0}$
& \cellcolor{orange!22} $2.69{\times}10^{1}$
& \cellcolor{orange!22} $3.65{\times}10^{1}$ \\[2pt]

& GNN
& \cellcolor{red!22} $6.68{\times}10^{1}$
& \cellcolor{red!22} $4.93{\times}10^{2}$
& \cellcolor{red!22} $5.72{\times}10^{2}$ \\
\midrule

\multirow{4}{*}{\textbf{LV9 258-bus}}
& HCP-PINN
& \cellcolor{green!30} $\mathbf{2.72{\times}10^{-13}}$
& \cellcolor{green!30} $\mathbf{3.27{\times}10^{-12}}$
& \cellcolor{green!30} $\mathbf{3.18{\times}10^{-11}}$ \\[2pt]

& WLS
& \cellcolor{green!12} $9.98{\times}10^{-4}$
& \cellcolor{green!12} $7.00{\times}10^{-3}$
& \cellcolor{green!12} $7.34{\times}10^{-3}$ \\[2pt]

& PINN
& \cellcolor{orange!22} $6.38{\times}10^{0}$
& \cellcolor{orange!22} $1.05{\times}10^{2}$
& \cellcolor{orange!22} $1.34{\times}10^{2}$ \\[2pt]

& GNN
& \cellcolor{red!22} $6.39{\times}10^{1}$
& \cellcolor{red!22} $9.19{\times}10^{2}$
& \cellcolor{red!22} $1.14{\times}10^{3}$ \\

\bottomrule
\end{tabular}
\end{table}

\subsection{Constraint Satisfaction and Physical Consistency}
This analysis evaluates how effectively HCP-PINN satisfies the equality constraints imposed through the \(\mathcal{S}_{\mathrm{MAP}}\) layer. In particular, nodal power balance constraints are enforced at zero-injection buses. Since these buses have no connected load or generation, their net active and reactive power injections are known exactly to be zero, providing additional noise-free physical information that can be incorporated as virtual measurements. Evaluating the corresponding power balance residuals therefore provides a direct measure of constraint satisfaction in the estimated states.

Table~\ref{tab:zi_pb} evaluates satisfaction of equality constraints for the LV2-43-bus and LV9-258-bus networks. The LV2-43-bus feeder contains 35 zero-injection buses, while the LV9-258-bus feeder contains 134 zero-injection buses. At zero-injection buses, HCP-PINN achieves zero residuals on both networks, substantially below unconstrained WLS, PINN, and GNN. The mean residuals are \(2.51\times10^{-11}\) and \(2.72\times10^{-13}\), while the maximum residuals remain below \(1.83\times10^{-9}\) and
\(3.18\times10^{-11}\), respectively, confirming that the constraints are
satisfied to high numerical precision even for the worst-case samples.   

\begin{figure}[!t]
  \centering
  \begin{subfigure}{0.48\linewidth}
    \centering
    \includegraphics[width=\linewidth,trim=0pt 0pt 5pt 3pt,clip]
    {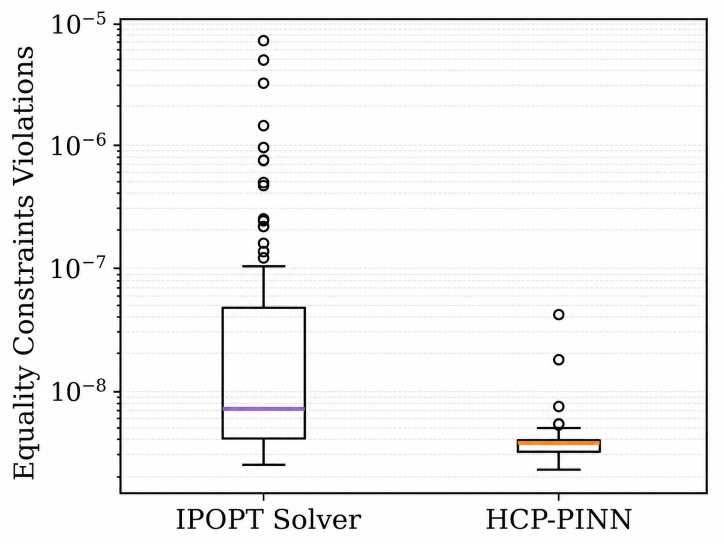}
    \caption{Constraint Violations}
    \label{fig:constraint_satisfaction}
  \end{subfigure}
  \hfill
  \begin{subfigure}{0.50\linewidth}
    \centering
    \includegraphics[width=.950\linewidth,trim=10pt 8pt 5pt 10pt,clip]
    {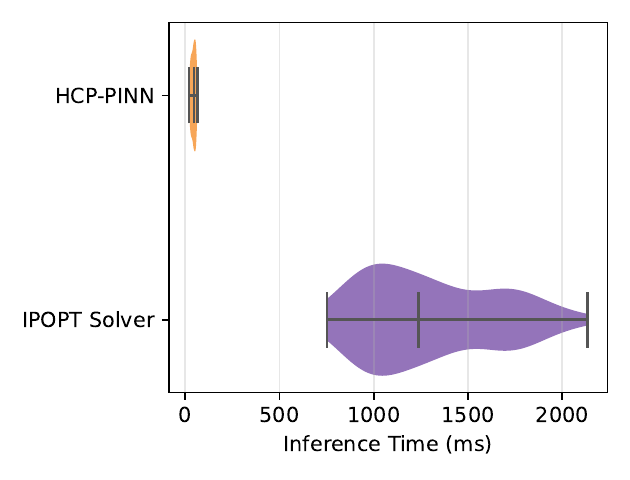}
    \caption{Time per sample (non-batch)}
    \label{fig:inference_time_solve}
  \end{subfigure}

  \caption{Constraint satisfaction and computational performance of HCP-PINN compared with IPOPT. (a) Distribution of equality-constraint violations. (b) Per-sample inference time.}
  \label{fig:objective_gap}
\end{figure}

Furthermore, HCP-PINN is compared with the constrained optimization solver \texttt{IPOPT}, a primal--dual interior-point method, using an equality-constrained WLS objective for the three-phase 258-bus feeder under partial observability. This comparison evaluates whether HCP-PINN can reproduce the solution quality and constraint satisfaction of a conventional optimization solver while providing a more computationally efficient learned alternative for constrained DSSE. The results show that HCP-PINN closely reproduces the constrained \texttt{IPOPT}-based SE solution, with a negligible objective gap and comparable voltage magnitude, phase angle, and line loading accuracy. For equality-constraint satisfaction, HCP-PINN achieves power balance residuals of \(2.72\times10^{-13}\) and \(2.64\times10^{-13}\), respectively, compared with \(1.09\times10^{-11}\) and \(8.43\times10^{-12}\) for \texttt{IPOPT}. These results confirm satisfaction of the equality constraints by both methods to high precision, with HCP-PINN yielding slightly smaller residuals, as also shown in Fig.~\ref{fig:constraint_satisfaction}.

Despite this comparable solution quality and physical consistency, HCP-PINN
substantially reduces computational cost. As shown in Fig.~\ref{fig:inference_time_solve}, the mean inference time decreases from \(1427\)~ms for \texttt{IPOPT} to approximately
\(57\)~ms per sample for HCP-PINN, corresponding to a speed-up of about
\(25\times\). Inference times were measured at batch size 1 using a single CPU thread on an Intel Core i7-12850HX workstation with 32~GB RAM, ensuring a fair per-sample comparison with the sequential \texttt{IPOPT} solver implemented through \texttt{Pyomo} library \cite{hart2011pyomo}. HCP-PINN can further exploit GPU-based batched inference for additional speed-up.

\subsection{Uncertainty Quantification}
In HCP-PINN, the FGNN provides a learned  prior, which is combined with the measurement likelihood within the \(\mathcal{S}_{\mathrm{MAP}}\) layer to obtain the final state estimate and associated predictive uncertainty. Because the exact posterior is generally intractable, it is locally approximated around the MAP solution using curvature information derived from a first-order linearization of the measurement model. This yields a sampling-free approximation of the posterior covariance, which is used to quantify uncertainty. The quality of the uncertainty estimates is evaluated using the CRPS metric, which assesses both calibration and sharpness of the predictive distribution, together with empirical coverage to evaluate the reliability of the prediction intervals. A CRPS value of zero corresponds to a perfect probabilistic prediction, with lower values indicating better predictive performance.
\begin{figure}[!t]
  \centering
  \begin{subfigure}[t]{0.49\linewidth}
    \centering
    \includegraphics[width=\linewidth,trim=0pt -20pt 0pt 23pt,clip]{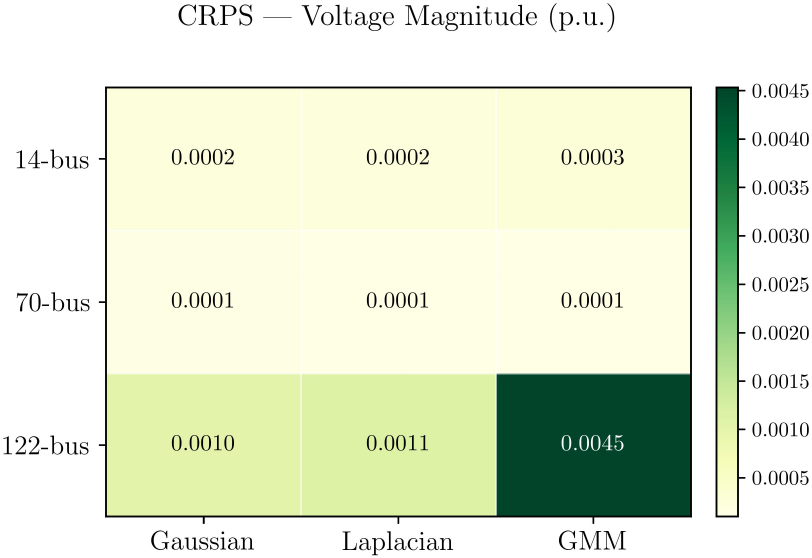}
    \caption{CRPS score.}
    \label{fig:crps_score}
  \end{subfigure}\hfill
  \begin{subfigure}[t]{0.46\linewidth}
    \centering
    \includegraphics[width=\linewidth,trim=0pt 0pt 0pt 25pt,clip]{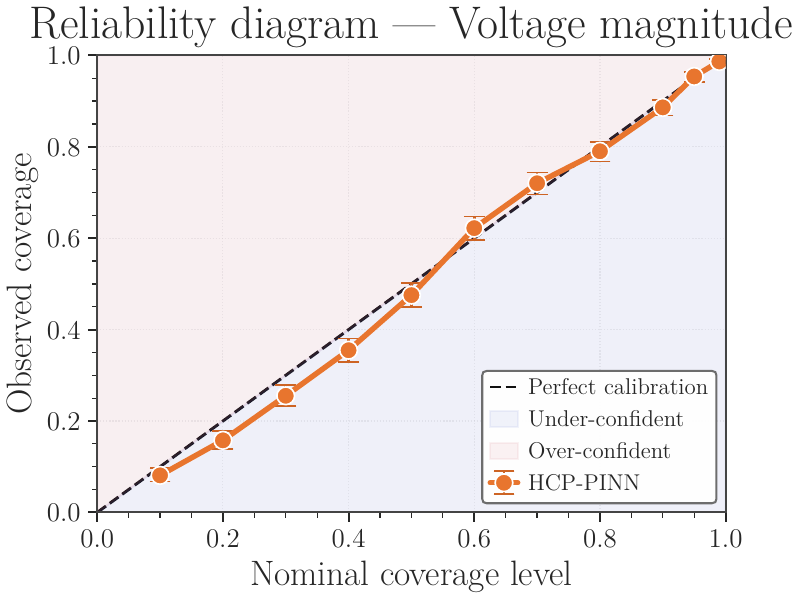}
    \caption{Reliability diagram.}
    \label{fig:reliability_diagram}
  \end{subfigure}
\caption{Uncertainty quantification metrics for voltage magnitude.
(a) CRPS scores across test networks and noise models, where lower values indicate better probabilistic accuracy.
(b) Reliability diagram comparing the nominal (expected) prediction-interval coverage with the observed empirical coverage; the dashed diagonal line represents perfect calibration.}
\label{fig:uq_metrics_vm}
\end{figure}
As shown in Fig.~\ref{fig:crps_score}, HCP-PINN achieves consistently low CRPS values of order \(10^{-4}\) for the 14-bus and 70-bus networks across all noise models. For the 122-bus network, the CRPS remains around $0.0010-0.0011$ under Gaussian and Laplacian noise but increases to $0.0045$ under GMM noise, indicating that multimodal uncertainty is more difficult to capture in the larger and more sparsely observed networks. Even under this more challenging setting, HCP-PINN maintains reasonable probabilistic accuracy.  

Fig.~\ref{fig:reliability_diagram} assesses the calibration of the predicted uncertainty by comparing the expected coverage of each prediction interval with the fraction of ground-truth values that actually fall within that interval. Perfect calibration is represented by the diagonal line, where the expected and observed coverages are equal. The HCP-PINN curve remains close to this diagonal across the evaluated confidence levels, indicating that its predicted uncertainty intervals generally contain the true states at the expected frequency.

\subsection{Sensitivity to Network Parameter Errors and Grid Upgrades}

In practice, utility grid models are constructed from geographical information systems (GIS), asset registers, and metering databases, which may contain incomplete, inaccurate, or outdated information that can compromise DSSE accuracy~\cite{gethDataQualityChallenges2023a}. This case study therefore evaluates the sensitivity of HCP-PINN to grid model inaccuracies and its generalization to future feeder upgrades. The analysis focuses on two common sources of grid model uncertainty: errors in line impedance parameters and incorrect phase assignments of single-phase customers. The underlying feeder connectivity is assumed to be known, as utilities generally have better knowledge of the main feeder topology, whereas line parameters and customer phase connectivity are more prone to inaccuracies. The considered test cases are summarized in Table~\ref{tab:robustness_cases}. 

\begin{table}[!t]
\centering
\caption{Test Cases for evaluating the robustness and generalization of the proposed DSSE framework on the three-phase unbalanced LV2-43bus network.}
\label{tab:robustness_cases}

\renewcommand{\arraystretch}{1.25}

\begin{tabular}{
@{}
>{\raggedright\arraybackslash}p{0.29\linewidth}
>{\raggedright\arraybackslash}p{0.70\linewidth}
@{}
}
\hline
\textbf{Test Case} & \textbf{Description} \\
\hline

\textbf{Baseline} &
Model evaluation under the correct network model and nominal operating conditions. \\

\textbf{Line Impedance errors} &
\textbf{Z5}, \textbf{Z10}, and \textbf{Z20}: 5\%, 10\%, and 20\% perturbations in line resistance $R$ and reactance $X$, respectively. \\

\textbf{Phase connectivity errors} &
\textbf{Ph1}, \textbf{Ph3}, and \textbf{Ph5}: incorrect phase assigned to 1, 3, and 5 different customers, respectively. \\

\textbf{New loads} &
\textbf{nL1}, \textbf{nL2}, and \textbf{nL3}: scenarios with 1, 2, and 3 newly introduced loads in the network, respectively. \\

\textbf{New PVs} &
\textbf{nPV1}, \textbf{nPV2}, and \textbf{nPV3}: scenarios with 1, 2, and 3 newly added PV units in the network, respectively. \\

\hline
\multicolumn{2}{@{}p{0.97\linewidth}@{}}{
\footnotesize\textit{Note:} nL/nPV locations are randomly chosen among zero-injection buses and may be connected to any bus phase.
} \\
\end{tabular}
\end{table}

Voltage magnitude RMSE is used to quantify the impact of these parameter errors, with 480 independent operating scenarios evaluated for each test case on the LV2-43bus feeder. As a reference, HCP-PINN is first evaluated using the correct network model under nominal operating conditions. The resulting median RMSE values are $2.95\times10^{-4}$ p.u. (0.074 V), $1.86\times10^{-4}$ p.u. (0.046 V), and $6.50\times10^{-4}$ p.u. (0.163 V) for phases A, B, and C, respectively.

\begin{figure*}[!t]
    \centering

    \begin{subfigure}[t]{0.49\textwidth}
        \centering
        \includegraphics[width=\linewidth,trim={11 0 0 0},clip]
        {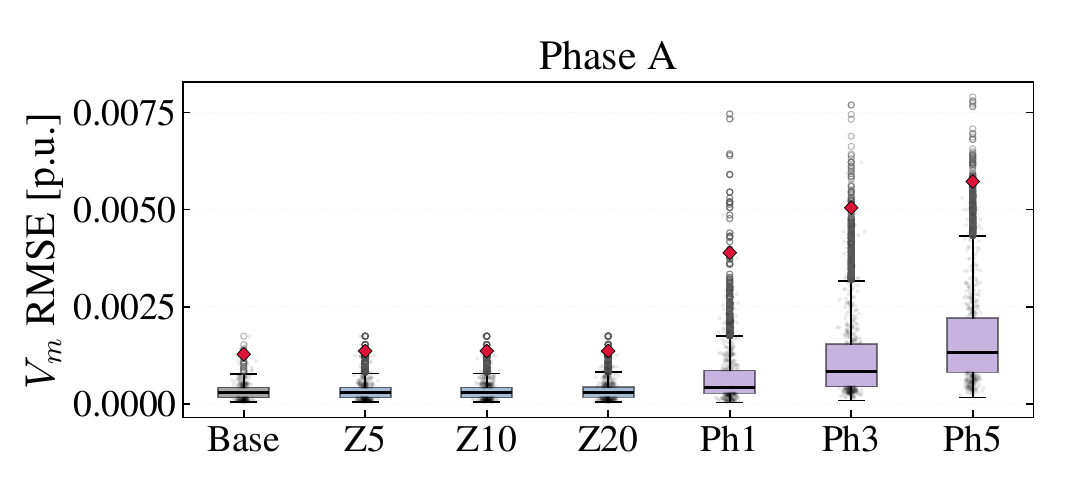}
        \label{fig:a}
    \end{subfigure}
    \hfill
    \begin{subfigure}[t]{0.49\textwidth}
        \centering
        \includegraphics[width=\linewidth,trim={15 0 0 0},clip]
        {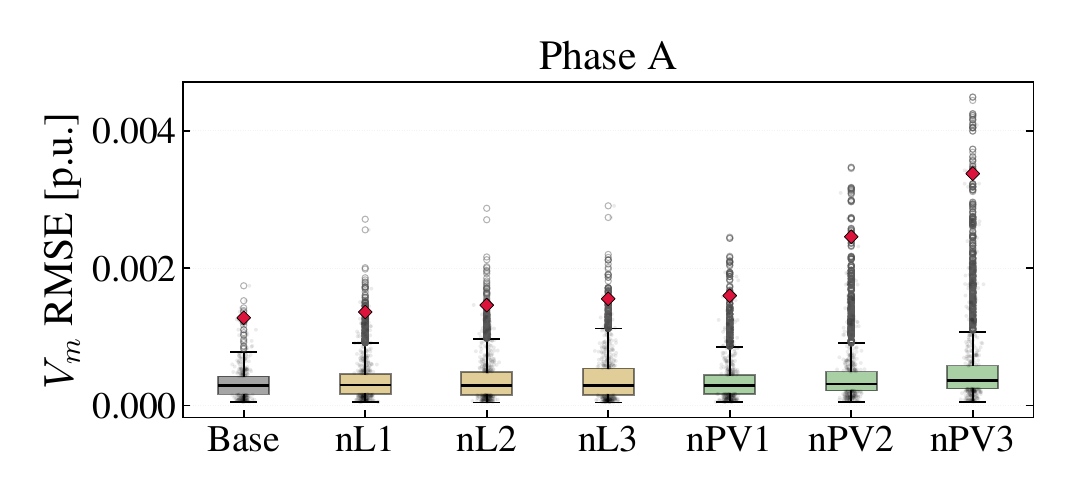}
        \label{fig:b}
    \end{subfigure}

    \begin{subfigure}[t]{0.49\textwidth}
        \centering
        \includegraphics[width=\linewidth,trim={15 0 0 0},clip]
        {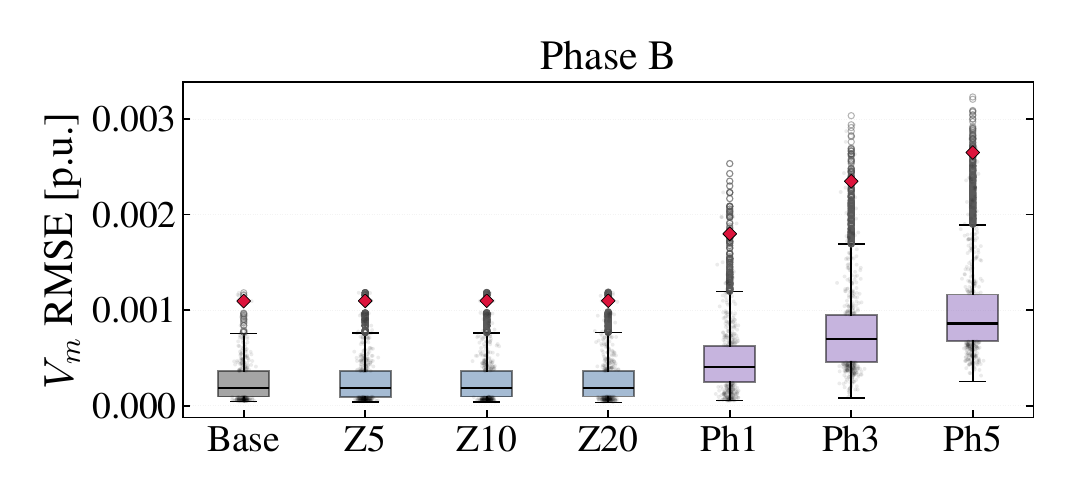}
        \label{fig:c}
    \end{subfigure}
    \hfill
    \begin{subfigure}[t]{0.49\textwidth}
        \centering
        \includegraphics[width=\linewidth,trim={15 0 0 0},clip]
        {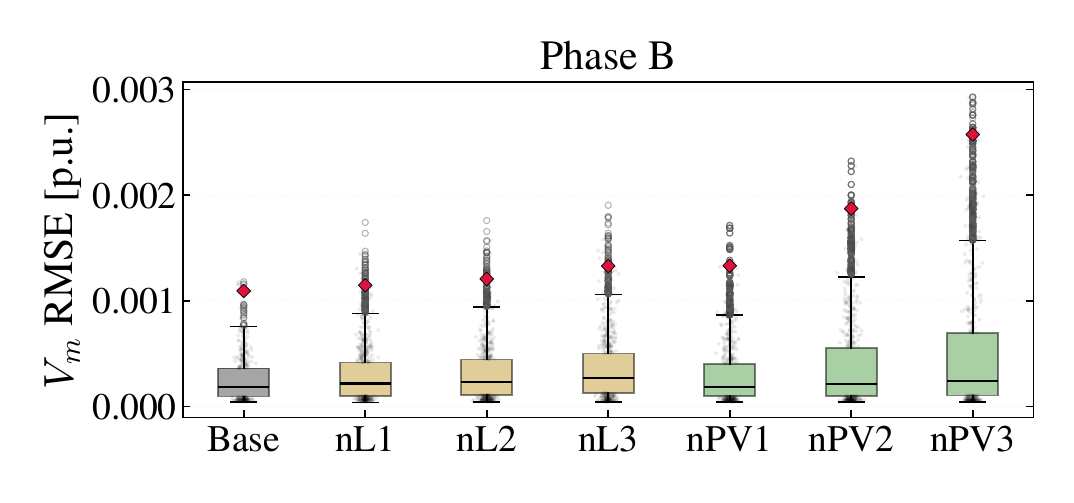}
        \label{fig:d}
    \end{subfigure}

    \begin{subfigure}[t]{0.49\textwidth}
        \centering
        \includegraphics[width=\linewidth,trim={15 0 0 0},clip]
        {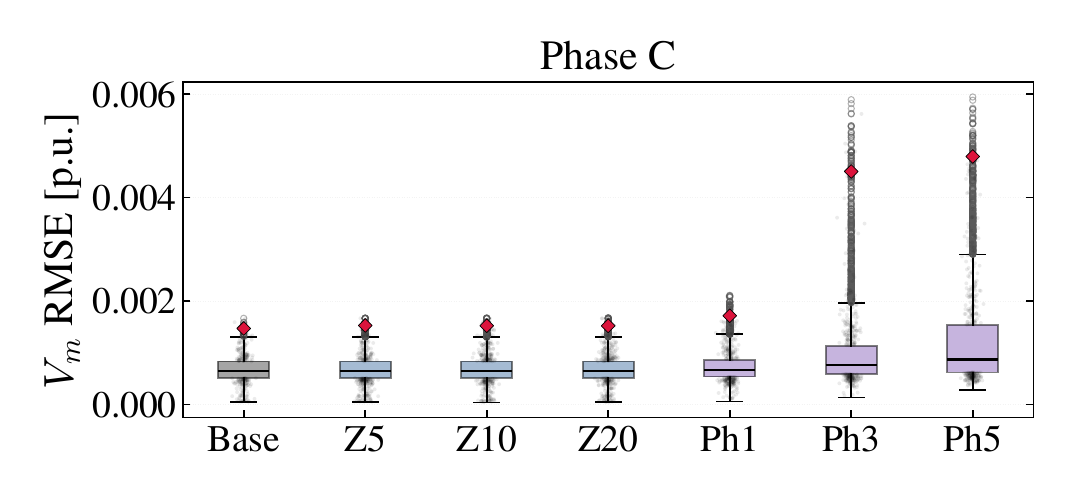}
        \captionsetup{labelformat=empty}
        \caption{\textbf{Grid data errors}}
        \label{fig:e}
    \end{subfigure}
    \hfill
    \begin{subfigure}[t]{0.49\textwidth}
        \centering
        \includegraphics[width=\linewidth,trim={15 0 0 0},clip]
        {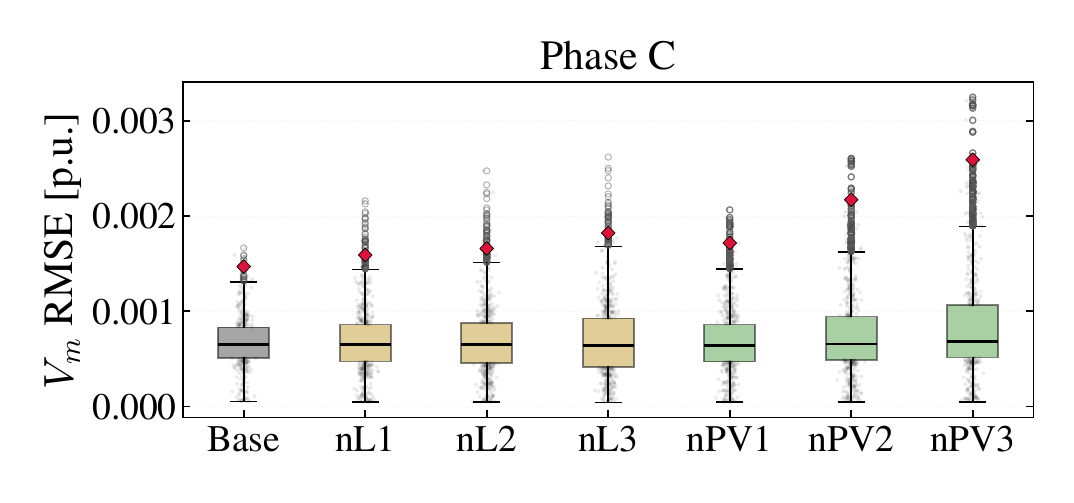}
        \captionsetup{labelformat=empty}
        \caption{\textbf{Grid upgrades}}
        \label{fig:f}
    \end{subfigure}

    \caption{voltage magnitude estimation errors on the LV2-48bus feeder under network-model errors and grid upgrades. Left: line impedance errors (Z5--Z20) and phase connectivity errors (Ph1, Ph3, Ph5). Right: new loads (nL1--nL3) and PV units (nPV1--nPV3) added at zero-injection buses. Distributions are summarized by the median and interquartile range; red markers denote the $99^{th}$ percentile errors.}
    \label{fig:sensitivity_analysis}
\end{figure*}

The first test case examines sensitivity to line-impedance uncertainty. Independent per-line perturbations of $\pm5\%$, $\pm10\%$, and $\pm20\%$ are applied to both resistance and reactance values (Z5--Z20). Across all perturbation levels, the median voltage magnitude RMSE remains below $6.50\times10^{-4}$ p.u. (0.163 V) for all phases as shown in Fig. \ref{fig:sensitivity_analysis}. Even for the largest perturbation (Z20), the increase relative to the baseline is negligible, and the overall error distributions remain largely unchanged.

The second network model error is the wrong phase assignment of single phase customers, which results in the dominant degradation in the model accuracy, as phase assignment errors modify the structural mapping between measurements/injections and the network equations. As the number of incorrect phase assignments increases, the maximum median \(V_m\) RMSE across all phases reaches \(1.32\times10^{-3}\) p.u. (0.33 V) at Ph5, corresponding to an increase of up to approximately \(4.6\times\) relative to the Base case as illustrated in Fig. \ref{fig:sensitivity_analysis}. The tail errors also increase, with the 99th-percentile RMSE reaching 1.43 V. This sensitivity is accentuated by the characteristics of the LV2-43bus feeder, which contains only seven single-phase customers; thus, Ph5 corresponds to incorrect phase assignments for five of the seven customers. The impact also varies across phases. Nevertheless, the median estimation error remains below 1 V, indicating a pronounced relative degradation rather than a loss of overall estimation capability. The voltage magnitude estimates remain comparatively accurate and robust because the voltage states are still constrained by the available measurements, while the phase-assignment mismatch is reflected more strongly in the form of larger power balance residuals.

Beyond network-model errors, the analysis considers plausible feeder changes after model deployment, particularly the addition of new residential loads and distributed PV systems at previously inactive zero-injection buses. Overall, these unseen grid upgrades cause only small changes in the median \(V_m\) RMSE, while the upper tail increases as more sites are added. For the newly added load cases (nL1--nL3), the median RMSE remains close to the Base case across all phases, with the largest relative increase of \(1.45\times\) observed on Phase B for nL3. The 99th-percentile RMSE remains below 0.46 V, indicating limited degradation when new residential loads are introduced at buses not seen during training. For the new PV cases (nPV1--nPV3), the median \(V_m\) RMSE also remains close to the Base case, reaching at most 0.17 V, whereas the upper tail increases more noticeably. On Phase A, the 99th-percentile RMSE rises from 0.32 V in the Base case to 0.40, 0.62, and 0.84 V for nPV1--nPV3, respectively. Although this tail degradation is larger than for the new load cases, the 99th-percentile error remains below 1 V.

As a final remark, defining a universal acceptance threshold for the proposed HCP-PINN under network parameter errors or the addition of new prosumers remains challenging. Based on the empirical results and related LV DSSE studies, voltage magnitude errors of approximately 1 V may be considered a practical upper bound, while values around 0.5 V or lower represent a desirable accuracy level. These values should be viewed as empirical targets rather than strict acceptance criteria.

\subsection{Generalization to Topological Changes and Operating Point Shifts}
Distribution grids frequently experience topological changes due to switching operations associated with load balancing, planned maintenance, and fault isolation or restoration. For conventional numerical methods, each topology change requires updating the network model and re-solving the estimation problem for the new configuration, which can be time-consuming. Learning-based DSSE methods offer faster inference after training; however, they must generalize beyond a fixed topology to modified network structures without requiring model retraining. To evaluate this generalization behavior, reconfigured topology instances of a 14-bus network are generated by applying \(k\) number of random switch changes, where each change opens one line and closes another tie line while preserving radial operation. For each value of \(k\), up to four distinct radial topological configurations are generated, and DSSE inference is performed using GNN and HCP-PINN models trained only on the original topology. 

\begin{figure}[!t]
  \centering

  \begin{subfigure}[t]{0.50\linewidth}
    \centering
    \includegraphics[width=\linewidth,trim=0pt 0pt 0pt 12pt,clip]
    {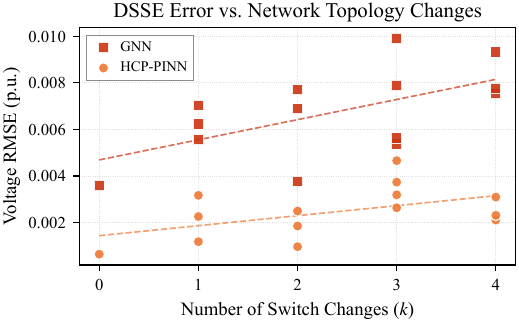}
    \caption{Topological changes.}
    \label{fig:topology_invariance_vm_rmse}
  \end{subfigure}
  \hfill
  \begin{subfigure}[t]{0.47\linewidth}
    \centering
    \includegraphics[width=\linewidth,trim=0pt 0pt 0pt 0pt,clip]
    {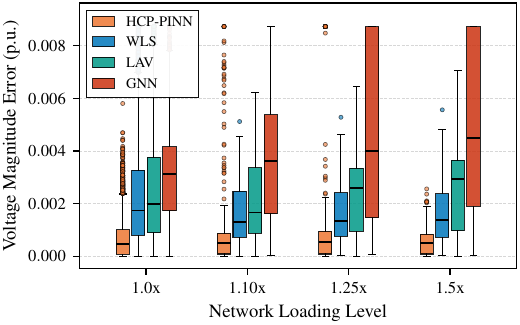}
    \caption{Operating-point shifts.}
    \label{fig:operating_point_shift_vm_rmse}
  \end{subfigure}

  \caption{Generalization performance under (a) network reconfigurations and
  (b) operating-point shifts, evaluated using voltage magnitude RMSE.}
  \label{fig:generalization_vm_rmse}
\end{figure}

Fig.~\ref{fig:topology_invariance_vm_rmse} illustrates the voltage magnitude RMSE as a function of the number of switch changes. The voltage estimation error increases with \(k\) because each reconfiguration modifies feeder connectivity and changes the power flows. This effect is more pronounced for the standard GNN because topological reconfiguration changes the neighborhood structure underlying its learned message-passing representations, making them less reliable as the topology deviates from the training graph. In contrast, HCP-PINN consistently achieves lower RMSE and exhibits less performance degradation across all topological reconfigurations. This improved generalization can be attributed to the factor graph formulation, which captures topology through localized variable--factor node interactions aligned with the power flow equations, and is complemented by a physics-constrained estimation layer that enforces consistency with the reconfigured network model. Consequently, HCP-PINN remains topology-aware while exploiting transferable physical relations among measurements, states, and constraints across different network configurations. 
%

Beyond topological variations, the generalization capability of HCP-PINN is further evaluated under operating point shifts by scaling the customer loads with factors $\alpha \in {1.00, 1.10, 1.25, 1.50}$. Such scaling represents realistic increases in existing customers demand due to electrification, such as the future installation of EV chargers and heat pumps by existing customers. This experiment assesses whether HCP-PINN can maintain accurate DSSE performance as the operating point shifts away from the nominal training regime toward more heavily loaded conditions. Fig.~\ref{fig:operating_point_shift_vm_rmse} shows the distribution of voltage magnitude errors under increasing network loading for HCP-PINN, WLS, LAV, and the standard GNN on the CIGRE 14-bus network under partial observability and normal measurement noise. As the loading level increases, HCP-PINN maintains a consistently low median error and a narrow interquartile range across all demand growth levels. This indicates that the proposed HCP-PINN provides stable and accurate voltage magnitude estimates under moderate-to-extreme network loading variations on existing customers buses. Numerical WLS and LAV remain relatively stable as $\alpha$ increases, but their errors are consistently much larger than those of HCP-PINN. This shows that HCP-PINN combines robustness to loading variations with substantially higher estimation accuracy.

Additionally, the purely data-driven GNN shows substantially larger errors and increasing dispersion at higher loading levels, indicating its limited extrapolation capability beyond the training regime. Although it leverages network topology, the absence of explicit physical constraints limits its generalization under shifted operating conditions, particularly under limited measurement coverage. HCP-PINN does not merely learn correlations tied to the loading patterns observed during training; instead, it learns physically meaningful relationships that remain effective under substantially higher loading conditions.

\begin{table}[!t]
\centering
\caption{Sequential ablation study of HCP-PINN on LV2-43bus, in which components are successively removed from the full model to quantify their contribution to estimation accuracy.}
\label{tab:hcppinn_ablation}
\small
\setlength{\tabcolsep}{8.5pt}
\renewcommand{\arraystretch}{1.2}

\begin{tabularx}{\linewidth}{
@{}
>{\raggedright\arraybackslash}X
cccc
@{}
}
\toprule
\textbf{HCP-PINN variant}
& \makecell{\textbf{\(\mathsf{V}_m\)}\\\textbf{RMSE}}
& \makecell{\textbf{\(\mathsf{V}_a\)}\\\textbf{RMSE}}
& \makecell{\textbf{Loading}\\\textbf{RMSE (\%)}}
& \makecell{\textbf{PB}\\\textbf{residuals}} \\
\midrule

HCP-PINN Complete Model
& \textbf{0.00080}
& \textbf{0.000902}
& \textbf{0.60}
& \textbf{0.00126} \\

Without equality constraints
& 0.00168 & 0.00205 & 0.70 & 0.00136 \\

Without surrogate backward rule
& 0.00181 & 0.00221 & 0.73 & 0.00138 \\

Without probabilistic FGNN prior
& 0.00203 & 0.00213 & 0.81 & 0.00152 \\

Without \(\mathcal{S}_{\mathrm{MAP}}\) layer (FGNN only)
& 0.00465 & 0.00218 & 1.95 & 0.00202 \\

\bottomrule
\end{tabularx}
\end{table}

\subsection{Ablation Study of the HCP-PINN}
Table~\ref{tab:hcppinn_ablation} summarizes the contribution of the individual HCP-PINN components through an ablation study, starting from the full model and successively removing the hard equality constraints, geometric surrogate
backward rule, probabilistic FGNN prior, and finally the
\(\mathcal{S}_{\mathrm{MAP}}\) estimation layer. The full HCP-PINN achieves the best overall performance across all electrical quantities. Removing the equality constraints increases the voltage magnitude and line loading errors and leads to larger power balance residuals, demonstrating the benefit of explicitly enforcing physical constraints. Further removing the geometric surrogate backward rule during model training causes an additional degradation in estimation accuracy. This indicates that the surrogate gradient enables end-to-end training through the estimation layer, providing a more informative learning signal to the FGNN with respect to the underlying constrained estimation objective.

When the probabilistic FGNN prior is replaced by using the FGNN prediction only as an initialization for \(\mathcal{S}_{\mathrm{MAP}}\), the voltage magnitude, line loading, and power balance accuracy deteriorate further, showing the benefit of retaining the learned state prior within the MAP formulation. The most pronounced degradation occurs when \(\mathcal{S}_{\mathrm{MAP}}\) itself is removed, reducing the architecture to the standalone FGNN only. Relative to this baseline, introducing the estimation layer alone reduces the voltage magnitude and line loading RMSE by approximately \(56\%\) and \(58\%\), respectively.
Overall, this ablation study shows that the constrained estimation
layer provides the largest improvement over the standalone FGNN, while the
probabilistic prior, surrogate backward rule, and hard equality constraints
provide additional and complementary gains in estimation accuracy and
physical consistency.

\section{Conclusions}
\label{sec6}
This paper proposed a novel hard-constrained, probabilistic, physics-informed neural network, termed HCP-PINN, for distribution system state estimation (DSSE) under non-Gaussian uncertainty in three-phase balanced and unbalanced networks. HCP-PINN formulates DSSE as a constrained probabilistic optimization problem on a factor graph and produces physically consistent state estimates through a differentiable estimation layer \(\mathcal{S}_{\mathrm{MAP}}\) that performs an optimality-seeking refinement step to explicitly enforce measurement consistency and equality constraints. 
The key innovation lies in integrating a probabilistic FGNN with the constrained $\mathcal{S}_{\mathrm{MAP}}$ layer, such that the learned state distribution is embedded directly into the estimation objective and jointly reconciled with the available measurements and physical network constraints. Unlike conventional projection- or warm-start-based approaches, the neural network therefore remains an integral part of the final state-estimation problem rather than serving only as an initializer. The framework supports flexible likelihood-based uncertainty modeling for continuous non-Gaussian measurement noise while providing well-calibrated probabilistic estimates for risk-aware decision making. Its factor graph representation solves DSSE on a bipartite graph that explicitly encodes relationships among state variables and measurements. This modular design captures complex higher-order dependencies and improves robustness to network reconfiguration and out-of-distribution operating conditions compared with standard GNNs.

To accelerate end-to-end training, a geometric surrogate-gradient backward rule was introduced to avoid computationally expensive unrolled or exact differentiation through the embedded \(\mathcal{S}_{\mathrm{MAP}}\) layer while providing stable gradient signals to the FGNN during backpropagation. Extensive case studies on balanced and unbalanced networks of different sizes show that HCP-PINN consistently outperforms conventional WLS and LAV estimators, penalty-based PINNs, and standard GNNs. Across different test cases, it reduces state-estimation errors by approximately 58\% compared with WLS/LAV and 85\% compared with soft-PINN/GNN baselines.HCP-PINN also achieves an approximately \(35\times\) inference speedup in non-batched settings and up to a \(125\times\) speedup with a batch size of 16 over numerical solvers on the larger 258-bus network, highlighting its computational efficiency and scalability. Furthermore, the experimental results indicate that HCP-PINN is less sensitive to network model errors and future grid upgrades. Overall, the results demonstrate robustness under sparse sensing, non-Gaussian noise, outliers, topology changes, and out-of-distribution scenarios, supporting its suitability for online monitoring of active distribution networks.

Future work may extend HCP-PINN to joint state estimation and system identification tasks, including line-parameter estimation and customer phase identification, enabling simultaneous network model correction and accurate state estimation.

\section*{Acknowledgements}
The authors acknowledge the financial support provided by the Flemish Institute for Technological Research (VITO), Belgium, and KU Leuven, Belgium, for the completion of this research work.

\bibliographystyle{IEEEtran}
\bibliography{references}

\end{document}